\pdfoutput=1
\documentclass[apj]{emulateapj}

\shorttitle{TESS Code}
\shortauthors{Duffell \& MacFadyen}

\begin{document}

\title{TESS: A Relativistic Hydrodynamics Code on a Moving Voronoi Mesh}

\author{Paul C. Duffell and Andrew I. MacFadyen}
\affil{Center for Cosmology and Particle Physics, New York University}
\email{pcd233@nyu.edu, macfadyen@nyu.edu}

\begin{abstract}

We have generalized a method for the numerical solution of hyperbolic systems of equations using a dynamic Voronoi tessellation of the computational domain.  The Voronoi tessellation is used to generate moving computational meshes for the solution of  multi-dimensional systems of conservation laws in finite-volume form.  The mesh generating points are free to move with arbitrary velocity, with the choice of zero velocity resulting in an Eulerian formulation.  Moving the points at the local fluid velocity makes the formulation effectively Lagrangian.  We have written the TESS code to solve the equations of compressible hydrodynamics and magnetohydrodynamics for both relativistic and non-relativistic fluids on a dynamic Voronoi mesh.  When run in Lagrangian mode, TESS is significantly less diffusive than fixed mesh codes and thus preserves contact discontinuities to high precision while also accurately capturing strong shock waves.  TESS is written for Cartesian, spherical and cylindrical coordinates and is modular so that auxilliary physics solvers are readily integrated into the TESS framework and so that the TESS framework can be readily adapted to solve general systems of equations.  We present results from a series of test problems to demonstrate the performance of TESS and to highlight some of the advantages of the dynamic tessellation method for solving challenging problems in astrophysical fluid dynamics.

\end{abstract}


\section{Introduction}
\label{sec:intro}

Many astrophysical gas dynamical systems involve the motion of
compressible fluid with relativistic velocity or energy density.
Techniques for the numerical solution of the equations governing the
multi-dimensional dynamics of relativistic fluids have been developed
greatly in recent years. Significant recent progress has been made in
grid-based methods where the fuid is described using Eulerian meshes
for hydrodynamics on both uniform meshes (see review by Marti \&
Muller, 2003, and references therein) and with adaptive mesh
refinement (AMR) \citep{duncan,ram,amrvac}, as well as using smoothed
particle hydrodynamics (Rosswog, 2010).  Extensions of these methods
have been implemented including for general relativistic
magnetohydrodynamics
\citep{font,harm,dh2003,echo,cosmos,pluto,wham,nagataki}
and for dynamic spacetimes \citep{anderson,cd08,etienne}.

In this paper we present a new method for
relativistic gas dynamics based on a dynamic Voronoi tessellation of
space.  The Voronoi tessellation generates a numerical mesh which
moves and distorts with an arbitrary velocity field depending on how
the mesh-generating points in the tessellation are moved. While holding
the mesh points fixed results in an Eulerian method, allowing them to
move with the local fluid velocity results in an effectively
Lagrangian method.  In this paper we present the TESS code which we
have written on a dynamic mesh generated by successive Voronoi
tesselations of the spatial domain. TESS is a numerical code and
general framework for both relativistic and non-relativistic
implementations of hydrodynamics and magnetohydrodynamics (MHD). The
strength of the method is to retain the advantages of conservation-law
formulation essential for accurate computation of relativistic gas
dynamics, while gaining the advantages of a Lagrangian scheme.  Of
particular importance, the mesh motion allows for contact discontinuities to
propagate without the excessive numerical diffusion often found in
fixed mesh computations.  The preservation of contact discontinuities
is of great importance for problems involving the development of fluid
instabilities and for reactive hydrodyamics where artifical diffusion
of reactant species can lead to unphysical solutions.  Using mesh
motion, TESS accurately solves the numericaly challenging case of
relativistic shear flows (see Section 3), a problem which
underresolved Eulerian simulations calculate incorrectly (Zhang \&
MacFadyen, 2006).

Lagrangian codes have had great success when employed in one
dimension, usually to treat problems with spherical symmetry.
Multidimensional problems, however, are more challenging for
Lagrangian codes, due to distortion of the computational mesh when
complexities in the flow develop.  \cite{n64} formulated a simple
strategy for dealing with mesh distortion in multiple dimensions,
which was to continuously remap the computational domain as the system
evolved, to prevent the mesh from becoming overly distorted.  Codes
employing this strategy are referred to as ``Arbitrary
Lagrangian-Eulerian'' (ALE) codes.  ALE codes solve the problem of mesh
distortion, but at the cost of diffusive remaps.

\cite{bp87} addressed the problem of mesh distortion by using a
tessellated computational domain.  The improvement was to employ the
Voronoi tessellation; a unique decomposition of space into polyhedra
based on a set of mesh-generating points (originally called ``fluid
markers'').  The advantage of such an approach is that the
mesh-generating points can move freely through the computational
domain and can be added or removed to enable adaptive refinement.  The Voronoi tessellation adjusts its shape and topology so
that the computational mesh does not become overly distorted.

\cite{w95} also made use of domain tessellation for mesh generation.
His ``signal method'' was a finite-volume method, partially inspired
by finite-element methods employed for solving problems with irregular
boundaries.  The method was first-order and conservative, and
employed a Delaunay triangulation of the computational domain.  The 
technique employed high-resolution shock-capturing techniques developed
over the last decades for grid-based Eulerian codes, while also having
the advantages that come from solving the fluid equations on a moving
mesh.  The ``grid cells'' in this case were Delaunay triangles, and
the main source of diffusion came about during changes in the mesh topology.
During such changes, the conserved quantities were redistributed
evenly among the affected triangles.  This process introduced diffusion
similar to that encountered in ALE codes during remapping of the mesh, but
because the diffusion was localized and only occurred during topology
changes, the accumulated error was reduced.

More recently, \cite{arepo} developed a second-order finite-volume
approach to solving Euler's equations using a Voronoi tessellation 
of the computational domain.  This method was implemented in the 
AREPO code, which has recently been applied to simulate star formation 
in a cosmological context \citep{arepo1}.  The advantage of using the
Voronoi tessellation instead of the Delaunay triangulation is that the
Voronoi cells do not suffer abrupt transformations during changes in
mesh topology so that re-mapping and fluid re-distribution are
uneccessary.  This was the first time a second-order finite-volume 
Lagrangian method of this nature was proposed.  Another important
property of this method Galilean invariance; the code's performance 
is independent of the velocity of the reference frame in which the 
calculation is performed.

In this paper we generalize these developments to the case of relativistic 
hydrodynamics.
We should not expect to retain all of the advantages found in the Newtonian 
case; in particular, the prospect of having a Lorentz-invariant formulation
is not to be expected, since in standard formulations mesh points 
are assumed to be simultaneous at each timestep.  

The paper is structured as follows: In \S 2, we describe the
formulation of the fluid equations, and the specific implementation in
the code.  In \S 3, we provide a series of test problems to determine
the convergence rate and to illustrate what sort of problems naturally
benefit from a Lagrangian approach.  In \S 4, we demonstrate the
code's usefulness in an astrophysical context, by looking at the
Rayleigh-Taylor instability in a decelerating relativistic shock.  In
\S 5, we summarize our results.

\section{Numerical Method}
\label{sec:form}

We first give a brief description of the Voronoi tessellation, before
explaining how it is used in solving the fluid equations.  A
tessellation is a decomposition of space into separate pieces, in our
case polygons or polyhedra.  For the moment, we restrict our attention
to two-dimensional tesselations, mainly because they are easier to
describe and to visualize.

\begin{figure*}
\epsscale{1.0}
\plotone{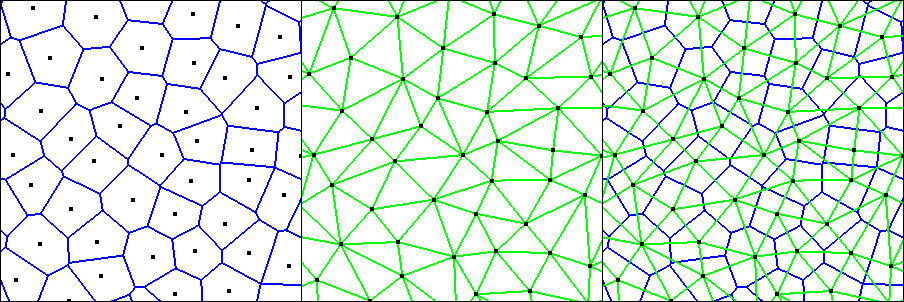}
\caption{ The Voronoi (left) and Delaunay (middle) tessellations applied to the same set of mesh points.  The right-most panel shows both tessellations superimposed, highlighting the duality between the two.  The Voronoi tessellation is the one employed in TESS, but the Delaunay triangulation is used to determine neighbors in an intermediate stage of the tessellation process.
\label{fig:tessellation} }
\end{figure*}

A Voronoi tessellation can be generated from some collection of points
(see Fig. \ref{fig:tessellation}).  Each mesh-generating point
(i.e. "mesh generator") will correspond to a polygon in the
tessellation.  The polygon associated with point P is defined to be
the set of all points which are closer to P than to any other mesh
generating point.  Thus, if an edge is adjacent to the two Voronoi
polygons of points P and Q, this edge lies in the line of points which
is equidistant to P and Q.  In general, we will refer to the polygons
as ``cells'' and the edges as ``faces''.  This terminology makes it
easier to generalize to arbitrary numbers of dimensions.
Additionally, we will speak of the ``volume'' of a cell, which in two
dimensions will mean area and in one dimension will mean length.
We will also refer to the ``area'' of a face, which in two dimensions
will mean length, and in one dimension will just be a constant number
which we set equal to unity.

An important related tessellation is the Delaunay triangulation.  Given a set of mesh generating points, one can generally form a triangulation with the mesh generators as vertices.  In a sense, the Delaunay tessellation is the ``nicest possible'' triangulation of the space, given by the following definition: for every triangle in the tessellation, the three vertices of the triangle all lie on some circle, C.  In order for this to be a proper Delaunay triangle, no other mesh generators may lie within the circle C.  This property almost uniquely defines the triangulation.  In degenerate cases, where four or more mesh generators lie on a common circle, the triangulation is ambiguous for these mesh generators.  Degenerate cases like this will not concern us greatly in this paper.

For a given set of mesh generators, the Delaunay tessellation is the topological dual to the Voronoi tessellation.  Every Delaunay edge corresponds to a Voronoi face, and every Voronoi vertex corresponds to a Delaunay triangle (in fact, the Voronoi vertex lies at the center of the circle generated by the three vertices of the Delaunay triangle).  This fact will help us in constructing the Voronoi tessellation, because the problem can be reduced to finding the equivalent Delaunay triangulation.  Since there is a straightforward test to check that a tessellation is Delaunay, this is typically the easiest way to find out which mesh generators are neighbors.

Before describing the details of how one might  generate a tessellation from a given set of points, it will be useful to write down the numerical formulation so that we know what information needs to be extracted from the tessellation procedure.  We shall find that our formulation applies generally to any hyperbolic system of conservation laws, regardless of whether the underlying equations are relativistic.

\subsection{Finite Volume Formulation}

The equations being solved will always have the form:
\begin{equation}
\partial_\mu f^\mu = S
\label{eqn:cons}
\end{equation}
\begin{equation}
f^\mu = \left( \begin{array}{c} U \\ \vec F \end{array} \right).
\end{equation}
In particular, for the case of Euler's equations:
\begin{equation}
U = \left( \begin{array}{c} \rho \\ S^i \\ E \end{array} \right)  F^j = \left( \begin{array}{c} \rho v^j \\ S^i v^j + P \delta^{ij} \\ E v^j + P v^j \end{array} \right)
\label{eqn:cons_nr}
\end{equation}
where $\rho$ is the density of the fluid, $\vec S$ is the momentum density, and $E$ is the energy density.  $P$ is the pressure, and $\vec v$ is the flow velocity.  
For simplicity we consider the case where there are no source terms, ${S = 0}$.
For the relativistic version:
\begin{equation}
U = \left( \begin{array}{c} \rho u^0 \\ \rho h u^i u^0 \\ \rho h u^0 u^0 - P - \rho u^0 \end{array} \right)  F^j = \left( \begin{array}{c} \rho u^j \\ \rho h u^i u^j + P \delta^{ij} \\ \rho h u^0 u^j - \rho u^j \end{array} \right).
\label{eqn:cons_rel}
\end{equation}
Where now ${u^\mu}$ is the four-velocity.  
\begin{equation}
\rho h = \rho + \epsilon + P
\end{equation}
and ${\epsilon}$ is the internal energy density, which can be found from the equation of state:
\begin{equation}
\epsilon = \epsilon ( \rho , P ).
\end{equation}
For the case of an adiabatic equation of state, we have
\begin{equation}
\epsilon = P/(\Gamma - 1)
\end{equation}
where $\Gamma$ is the adiabatic index of the fluid.  We consider general physical equations of state in the Appendix.  To derive the finite volume form of these equations, we shall set the source term to zero for brevity, but the generalization to a nonzero source term is straightforward.  For concreteness, we shall work in 2+1 dimensions, but everything said here easily generalizes to arbitrary numbers of dimensions.  Equation (\ref{eqn:cons}) does not depend on the spacetime metric a priori, and so we can write down a formulation independent of the metric.  For the following derivation we shall assume a Euclidean metric, rather than a Minkowski metric.

\begin{figure}
\epsscale{1.0}
\plotone{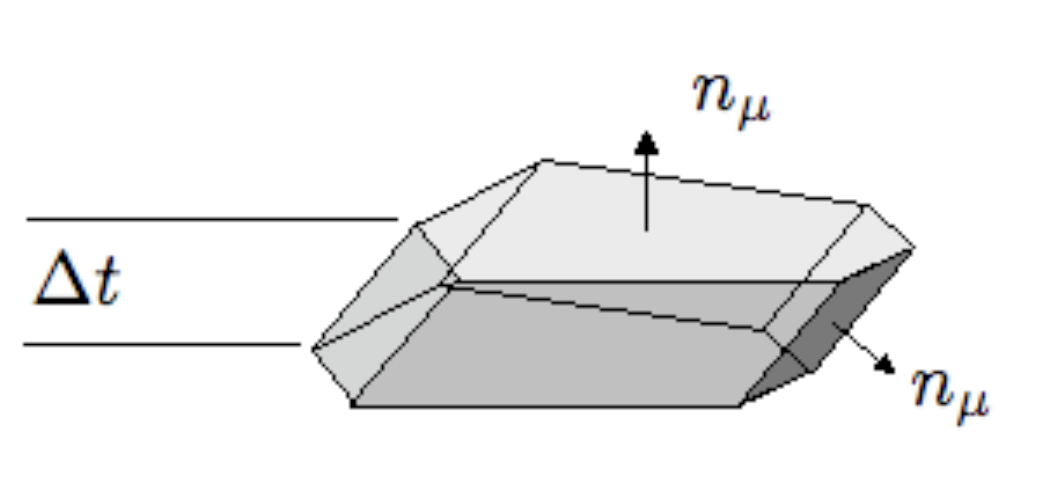}
\caption{ The solid polyhedron traced out by a Voronoi cell during one timestep.  The conservation law can be interpreted as "zero net flux" through all the faces of this spacetime polyhedron.
\label{fig:poly} }
\end{figure}

\begin{figure*}
\epsscale{1.0}
\plotone{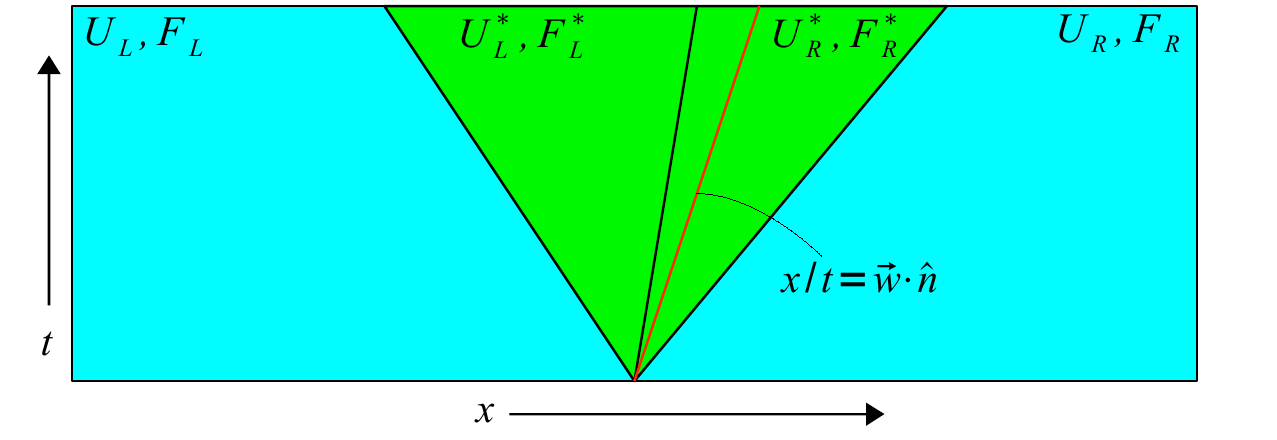}
\caption{ A space-time diagram of the Riemann problem with a moving face.  The HLLC solver assumes two constant states in the interior of the Riemann fan, separated by a contact discontinuity.  An eulerian code selects the state which houses the curve ${x/t = 0}$, whereas our formulation selects the state which houses ${x/t = \vec w \cdot \hat n}$ (in this example, the face lies in the region corresponding to ${U^*_R, F^*_R}$).  Note that if a face moves superluminally, our scheme reduces to upwinding.
\label{fig:rfan} }
\end{figure*}

We now look at the evolution of a Voronoi cell over one timestep.  It is assumed that the cell changes its shape and size by the linear motion of its faces, and that over a timestep it traces out a solid polyhedron in 2+1 dimensions, whose top and bottom are surfaces of constant time (see Fig. \ref{fig:poly}).  The shape is actually not quite a polyhedron, but this is an approximation we are making, which is valid in the limit of short timesteps.  In practice, we resolve this issue and others by taking a high-order Runge-Kutta timestep.  We integrate (\ref{eqn:cons}) over this polyhedral 2+1-dimensional volume P:
\begin{equation}
\int_{P} \partial_\mu f^\mu d^3x = 0.
\end{equation}
We can easily convert this to an integral over the two-dimensional boundary of this solid:
\begin{equation}
\int_{\partial P} f^\mu n_\mu d^2x = 0
\label{eqn:int}
\end{equation}
where ${n_\mu}$ is the (euclidean) unit normal to the boundary.  For the top and bottom of the polyhedron:
\begin{equation}
n_\mu = \left( \begin{array}{c} \pm 1 \\ \vec 0 \end{array} \right).
\end{equation}
The 2+1-dimensional unit normal to the other faces will be related to the 2-dimensional unit normal on a given timeslice, ${\hat n}$, but it will also have a component in the time dimension because the face is moving with some velocity, ${\vec w}$.  If we assume that the face is not changing its size or orientation as it moves (another assumption which is resolved by taking a high-order Runge-Kutta timestep), it is straightforward to check that the 2+1-dimensional unit normal will be
\begin{equation}
n_\mu = { 1 \over \sqrt{1 + (\vec w \cdot \hat n)^2}} \left( \begin{array}{c} - \vec w \cdot \hat n \\ \hat n \end{array} \right).
\end{equation}
Now we evaluate the integrals in (\ref{eqn:int}) by averaging the integrated quantities over spacetime.  In doing so, we need to know the 2+1 dimensional spacetime volume being integrated over.  For the top and bottom, this is straightforward:
\begin{equation}
\int_{Bottom} d^2x = dV^n, \int_{Top} d^2x = dV^{n+1}
\end{equation}
where dV is the cell volume at the beginning or end of the timestep.  For the other faces, it is easy to check that:
\begin{equation}
\int_{Face} d^2x = \sqrt{ 1 + (\vec w \cdot \hat n)^2 }dA \Delta t.
\end{equation}
Recall that ``dA'' is the face ``area'', so it refers to the length of a Voronoi edge.  Note that our factors of ${\sqrt{ 1 + (\vec w \cdot \hat n)^2 }}$ will end up cancelling, which is to be expected if our formulation is independent of the spacetime metric.  If we interpret ${U^n}$ as the cell-averaged value of U at timestep n, and let ${F_{ij}}$ denote the time-averaged flux through the face adjacent to cells i and j, and likewise use ${U_{ij}}$ to denote the time-and-area-averaged value of U on this same face, we get the following result:
\begin{eqnarray}
\begin{array}{c}
0 = \int_{\partial P} f^\mu n_\mu d^2x = \hspace{80pt} \\
\\
U^{n+1} dV^{n+1} - U^n dV^n + \Delta t \sum\limits_{cell j} ( F_{ij} - \vec w \cdot \hat n U_{ij} ) dA_j . \\
\end{array}
\end{eqnarray}
This gives us a simple prescription for how to evolve the conserved variables from one timestep to the next, assuming we know the time-averaged fluxes and conserved quantities on the faces, and the velocity of each face:
\begin{equation}
U^{n+1} dV^{n+1} = U^n dV^n - \Delta t \sum\limits_{cell j} ( F_{ij} - \vec w_{ij} \cdot \hat n U_{ij} ) dA_j .
\label{eqn:evolve}
\end{equation}
Of course, this result is not surprising at all.  The prescription merely tells us to add an advective term to the flux (${\vec F \rightarrow \vec F - \vec w U}$), and evolve things in a way analogous to the fixed-grid approach.  It is worth noting that our formulation does not depend on the physical content of the equations expressed by (\ref{eqn:cons}).  In particular, it does not matter whether the velocity ${\vec w}$ exceeds the speed of light.  In order for this to be possible, we must be careful about our physical interpretation for the mesh itself.  For example, it is not necessarily meaningful to speak of the ``rest frame'' of a cell or face.

\subsection{Riemann Solver}
\label{sec:riemann}

Equation (\ref{eqn:evolve}) approaches an exact result, in the limit of short timesteps.  This means that all of the numerical error due to the spatial discretization must stem from our estimates of the time-averaged values of the fluxes and conserved quantities on the faces of the Voronoi cells.  In TESS, we estimate these fluxes using an approximate Riemann solver.  This means we assume that, at the beginning of the timestep, there is a uniform state on either side of the face.  Generally speaking, a Riemann solver estimates the time-averaged flux through the face by evolving this constant-state initial condition, either exactly or approximately.

It turns out that getting the most out of the Lagrangian nature of our scheme is highly dependent on how we solve the Riemann problem on each cell face.  The Riemann solver in AREPO \citep{arepo} is effective because it is Galilean-invariant; the Riemann problem is solved in the rest frame of the face.  In our case we cannot assume it makes sense to even speak of a ``rest frame'' for a face, so we cannot have a method which is Galilean-invariant (this is to be expected anyway for relativistic problems).  Nonetheless, we can still have a code which preserves contact discontinuities to very high accuracy.  What we require is a Riemann solver which is exact for a pure advection problem.  The HLLC Riemann solver, appropriately employed, meets this requirement.  In the TESS code, we employ the HLLC solver for relativistic problems as described by \cite{hllc}.

A pure advection problem has constant pressure and velocity on both states, and two different densities.  The HLLC solver divides spacetime into four pieces: the original left and right state, and two interior states known as *L and *R, separated by a contact discontinuity.  For advection, we have the following:
\begin{equation}
\begin{array}{c}
\rho_{*L} = \rho_L \\
\rho_{*R} = \rho_R \\
P_{*L} = P_{*R} = P_{L} = P_{R} \\
v_{*L} = v_{*R} = v_{L} = v_{R} .
\end{array}
\end{equation}
This is the exact solution to the advection problem, and hence when using the HLLC solver, we can solve the advective Riemann problem to machine precision, as long as the correct starred state is chosen, i.e. the solver should choose the spacetime region which houses the path traced out by the face as it moves with velocity ${\vec w}$ (see Fig. \ref{fig:rfan}).  The values ${F^*}$ and ${U^*}$ found in this spacetime region are the values used in the update step given by equation (\ref{eqn:evolve}).  The reason that the HLLC solver is so crucial is that it accurately calculates advective fluxes.  Since the numerical error caused by the spatial discretization is entirely housed in estimating the time-averaged fluxes ${\vec F \cdot \hat n - \vec w \cdot \hat n U}$, and if the velocity of each face is very close to the fluid velocity, then for HLLC the advective fluxes will cancel, so that the numerical error for advective fluxes will be small.  In the particular case of pure advection, advective fluxes completely cancel, meaning that the advection problem is solved exactly to machine precision.  This property is extremely important for preserving contact discontinuities.

\subsection{Primitive Variable Solver}
\label{sec:roots}

The Riemann solver takes in two states corresponding to two adjacent Voronoi cells.  To use the information in a cell to solve the Riemann problem for a given face, we need the primitive variables (density, pressure, and four velocity) on either side of the face.  In order to find these primitive variables, we need to invert formulas (\ref{eqn:cons_nr}) or (\ref{eqn:cons_rel}) for the conserved variables on either side.  This can be done using a Newton-Raphson rootfinding scheme.  For relativistic hydrodynamics with an adiabatic equation of state, this solver is not difficult to write.  For an arbitrary equation of state, we incorporate the temperature as an additional variable which must be solved for.  In addition to an arbitrary equation of state, we want TESS to become a very general code capable of solving wide classes of problems, for example the equations of general relativistic magnetohydrodynamics (GRMHD).  This makes the Newton-Raphson step somewhat more complicated, but this has already been employed in TESS even though we have not yet employed hydrodynamic variables which would necessitate such a solver.  The GRMHD primitive variable solver is based on the method described by \cite{cons2prim} and uses a three-dimensional Newton-Raphson step, solving for the variables ${W = \gamma^2 \rho h}$, ${K = \gamma^2 - 1}$, and the temperature, $T$ (see the Appendix for more details).

\subsection{Reconstruction of Face-Centered Primitive Variables}
\label{sec:plm}
The root finding method described above determines the primitive variables (density, pressure, and four-velocity) at the cell centers.  To solve the Riemann problem on each face, we must extrapolate the primitive variables from the cell centers to the face centers.  If we assume all our variables to be piecewise constant, then we can assume they have the same value on the face as they do at the center of mass of a cell.  However, if we want accuracy higher than first order in space, we need to extrapolate the variable values based on the values in neighboring cells.  We use piecewise linear reconstruction, using the method derived by \cite{arepo} for calculating the variable gradients in each cell.  We repeat the results here.  Assume we have some variable, $\phi$, for which we would like to calculate the gradient at cell i based on the values at adjacent cells.  The formula we use is
\begin{equation}
\left< \vec \nabla \phi \right>_i = {1 \over V_i} \Sigma_j dA_j ( [\phi_j - \phi_i] {\vec c_{ij} \over r_{ij}} - {\phi_i + \phi_j \over 2}{\vec r_{ij} \over r_{ij}} ).
\end{equation} 
We use this gradient to extrapolate primitive variable values via:
\begin{equation}
\phi(\vec f_{ij}) = \phi_i + \left< \vec \nabla \phi \right>_i \cdot (\vec f_{ij} - \vec s_i).
\end{equation}
Here, ${\vec r}$ represents the location of a mesh generator, ${\vec s}$ represents the location of the center of mass of a cell, and ${\vec f_{ij}}$ is the center of mass of the face adjacent to cells i and j.  Note that primitive variables are defined at ${\vec s}$, not at the mesh generators.  This prescription would lead to a code which is second order in space, but it is well known that piecewise linear reconstruction can cause numerical oscialltions in the neighborhood of shocks.  To deal with this, we need to constrain the estimated gradients in the neighborhood of sharp discontinuities.  In other words, we need to construct a generalization of the slope limiters used in grid-based Eulerian codes.

AREPO uses a slope limiter which could be considered a generalization of minmod \citep{kur00}, but one which does not have the total variation diminishing (TVD) property.  As a result, this slope limiter caused mild oscillations in some calculations.  TVD is an especially important property for relativistic hydrodynamics, since oscillatory behavior in the conserved variables can potentially cause wild variation of the primitive variables, particularly in situations with large Lorentz factors.  To address this problem, we optionally employ an alternative slope limiter which is much more conservative.  AREPO uses the slope-limited gradient,

\begin{equation}
\left< \vec \nabla \phi \right>'_i = \alpha_i \left< \vec \nabla \phi \right>_i
\end{equation}
\begin{equation}
\alpha_i = min(1,\psi_{ij})
\end{equation}
\begin{equation}
\psi_{ij} = \left\{ \begin{array}
                  {l@{\quad:\quad}l}
                 (\psi_i^{max} - \psi_i)/\Delta \psi_{ij} & \Delta \psi_{ij} > 0  \\  
                 (\psi_i^{min} - \psi_i)/\Delta \psi_{ij} & \Delta \psi_{ij} < 0  \\
                 1 & \Delta \psi_{ij} = 0 
                  \end{array} \right.
\label{eqn:minmod}
\end{equation}
Our method is similar, but replaces (\ref{eqn:minmod}) with
\begin{equation}
\psi_{ij} = \left\{ \begin{array}
                  {l@{\quad:\quad}l}
                 max( \theta (\psi_j - \psi_i)/\Delta \psi_{ij} , 0) & \Delta \psi_{ij} > 0  \\  
                 max( \theta (\psi_j - \psi_i)/\Delta \psi_{ij} , 0) & \Delta \psi_{ij} < 0  \\
                 1 & \Delta \psi_{ij} = 0 
                  \end{array} \right.
\label{eqn:minmod2}
\end{equation}
where ${\theta}$ is generally set to unity, but reduced if a still more diffusive scheme is desired.  The slope limiter in (\ref{eqn:minmod2}) enforces monotonicity, though it is more diffusive than (\ref{eqn:minmod}).  In practice, we find it to be much more robust, so it has typically been employed in problems involving strong shocks.

\subsection{Time Integration}
\label{sec:time}
Our time integration is based on the method of lines, performing a Runge-Kutta timestep for the time evolution of the conserved variables, and for the motion of the mesh points.  For most problems, we use a third order Runge-Kutta timestep which is TVD (total variation diminishing,) and which updates both the values of the conserved variables, and the positions of the mesh generating points.  The timestep is Courant-limited:
\begin{equation}
\Delta t = C_{cfl} \cdot min( { R_i \over | \lambda^{max}_i | })
\end{equation}
${C_{cfl}}$ is the Courant factor, typically chosen between 0.2 and 0.5.  ${R_i}$ is the effective radius of a cell, ${R = \sqrt{dV/\pi}}$ in 2D.  ${\lambda^{max}_i}$ is the eigenvalue in cell i with the largest magnitude.  Currently the code is set up to operate with a single global timestep.  To take a timestep from ${U^n}$ to ${U^{n+1}}$ which is third-order in time, for example, we use the following prescription:
\begin{equation}
\begin{array}{rcl}
U^{(1)} & = & U^n + \Delta t L( U^n , \vec r^n ) \\
\vec r^{(1)} & = & \vec r^n + \Delta t \vec w^n\\
\\
U^{(2)} & = & {3 \over 4}U^n + {1 \over 4} U^{(1)} + {1 \over 4} \Delta t L( U^{(1)} , \vec r^{(1)} ) \\
\vec r^{(2)} & = & {3 \over 4} \vec r^n + {1 \over 4} \vec r^{(1)} + {1 \over 4} \Delta t \vec w^{(1)} \\
\\
U^{n+1} & = & {1 \over 3}U^n + {2 \over 3} U^{(2)} + {2 \over 3} \Delta t L( U^{(2)} , \vec r^{(2)} ) \\
\vec r^{n+1} & = & {1 \over 3} \vec r^n + {2 \over 3} \vec r^{(2)} + {2 \over 3} \Delta t \vec w^{(2)} \\
\end{array}
\end{equation}
Here, L is an operator representing the numerically integrated time derivative of U, and the variables ${U^{(1)}, U^{(2)}, \vec r^{(1)}}$, and ${\vec r^{(2)}}$ represent intermediate states in the time integration.

\subsection{The Voronoi Mesh}
\label{sec:mesh}

Equation (\ref{eqn:evolve}) tells us exactly what geometric information we need in order to evolve the conserved quantities.  We need to know the following about the Voronoi cells:

\begin{itemize} 
\item which cells are neighbors 
\item the volume of each cell 
\item the area of each face 
\item the velocity of each face 
\item the center of mass of each cell 
\item the center of mass of each face 
\end{itemize}

The last two elements of this list are necessary for the piecewise linear reconstruction of primitive variables.  We must determine how to extract all of this information given the positions and velocities of all the mesh generating points.  The velocities of mesh generators are freely specifiable, and we shall typically choose to set them equal to the local fluid velocity.

Before we determine this completely, we can show that all of the above can be calculated easily if we know which cells are neighbors, and if we also know the center of mass and area of each face.  In other words, when performing the Voronoi tessellation, this will be the only information we need to store.

Given a single mesh generator and its neighbors, it is straightforward to calculate its cell's volume, if the area of each face is known.  This is done by dividing the cell into pyramids (in 2D, triangles), each of whose tip is the mesh generating point, and whose base is a Voronoi face.  Then the volume of the cell is the sum of the volumes of all the pyramids:

\begin{equation}
dV_i = \Sigma_j V_{\Delta j}
\label{eqn:begingeo}
\end{equation}
The volume of a pyramid can be expressed generally in D-dimensions:
\begin{equation}
V_\Delta = (\mbox{area of base})(\mbox{height})/D
\end{equation}
Because the face is in the plane halfway between the two mesh generating points, the height should be half the distance between these points.
\begin{equation}
V_{\Delta j} = dA_{ij} (|r_{ij}|/2) /D
\end{equation}
\begin{equation}
dV_i = \Sigma_j {dA_j (|r_{ij}|/2) \over D}
\end{equation}
Similarly, the center of mass of a cell can be directly calculated from the area and center of mass of the faces.  We can use the weighted average over pyramids:
\begin{eqnarray}
\vec s_i = {1 \over dV_i} \Sigma_j V_{\Delta j} \vec s_{\Delta j} \\
= {1 \over dV_i} \Sigma_j {dA_j (|r_{ij}|/2) \over D} \vec s_{\Delta j} .
\end{eqnarray}
The center of mass of a pyramid also depends on the number of dimensions:
\begin{equation}
\vec s_{\Delta j} = {D \over D + 1} \vec f_{ij} + {1 \over D + 1} \vec r_i .
\end{equation}
Here, ${\vec f_{ij}}$ denotes the center of mass of the face adjacent to cells i and j.

Next, we need to determine the velocity of the faces.  It is assumed here that the mesh-generating points themselves have been given some velocity ${\vec w_i}$, typically the local fluid velocity (though corrections can be added to steer the cells in ways that make the mesh better-behaved).  \cite{arepo} showed that the velocity of a face can be calculated from the position and velocity of the mesh generating points and the center of mass of the face.  The result is the average of the velocity of the two adjacent mesh generators, plus a correction due to the fact that the center of mass of the face is not generally at the midpoint between the two mesh generating points, and so acquires a velocity due to rotation about this midpoint:
\begin{equation}
\vec w_F = (\vec w_L + \vec w_R)/2 + \vec w'
\end{equation}
\begin{equation}
\vec w' = (\vec w_L - \vec w_R) \cdot ( \vec f - (\vec r_L + \vec r_R)/2 ) { \vec r_R - \vec r_L  \over (\vec r_R - \vec r_L)^2 } .
\label{eqn:endgeo}
\end{equation}
Again, ${\vec f}$ is the center of mass of a face.  We can use equations (\ref{eqn:begingeo} - \ref{eqn:endgeo}) to pare down the list of information that we need to extract directly from the tessellation itself.  The tessellation procedure now only consists in determining the following:
\begin{itemize}
\item which cells are neighbors
\item the area of each face
\item the center of mass of each face
\end{itemize}
All relevant geometrical information can be easily extracted from this data.  This is advantageous because it means the tessellation can take up a relatively small amount of memory.  The tessellation procedure consists of generating a new set of faces each timestep, based on the locations of the mesh generators. 

In one dimension, the tessellation procedure is trivial; neighboring cells do not change, the face area is always set to unity, and the face center of mass is simply the midpoint between the two mesh generators.  The two-dimensional version turns out to be surprisingly simple, because we use the tessellation from the previous timestep to generate the new faces.  Since we know the neighbors of each cell on the previous timestep, we can use the neighbors of neighbors (``friends of friends'') of a cell as a list of candidates for the neighbors on the next timestep.  Because the length of each step in time is Courant-limited, the tessellation will not change significantly in one timestep, and hence this list of candidates is big enough for our purposes.  Optionally, we can choose to use ``neighbors of neighbors of neighbors'' but we have not found this to make a difference in any scenario we've encountered.  Already having this list of candidates simplifies the algorithm immensely, as in principle it can reduce to a brute force guess-and-check procedure using this small list of candidates.  In practice, the 2D algorithm is not totally brute force, but very simple nonetheless.

\begin{figure}
\epsscale{1.0}
\plotone{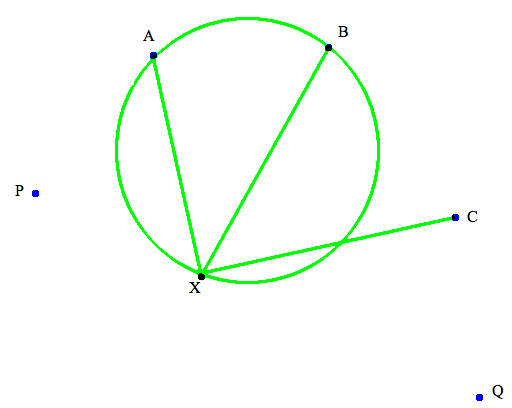}
\caption{ A single step of the tesselation process.  We decide whether to keep point B by asking whether point C is outside of the circle generated by X, A, and B.
\label{fig:tess1} }
\end{figure}

\begin{figure*}
\epsscale{1.0}
\plotone{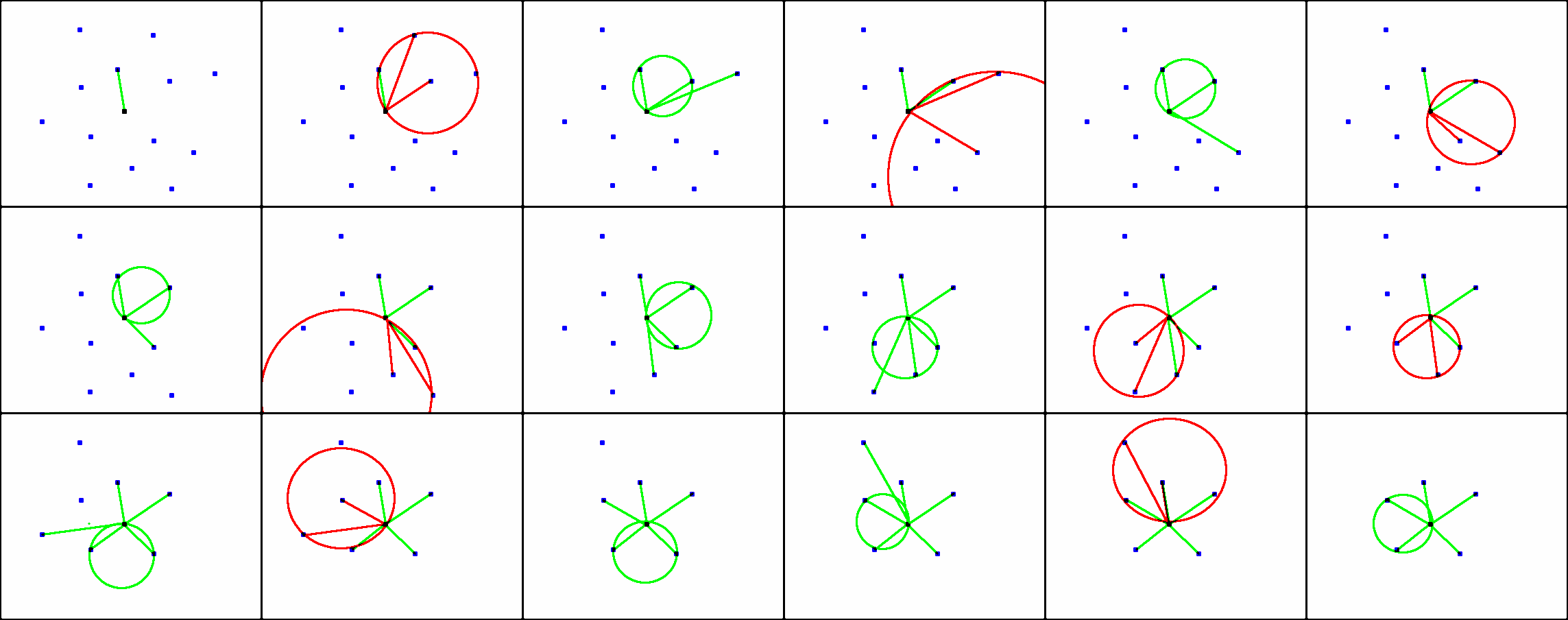}
\caption{ A demonstration of the tessellation algorithm in two dimensions.  Potential neighbors are sequentially tested and either accepted or rejected based on the criterion in Fig. \ref{fig:tess1}.  Red circles indicate when a candidate is being rejected.  This is just a sketch of the algorithm; circles are not accurately plotted.
\label{fig:tess2} }
\end{figure*}

\begin{figure}
\epsscale{1.0}
\plotone{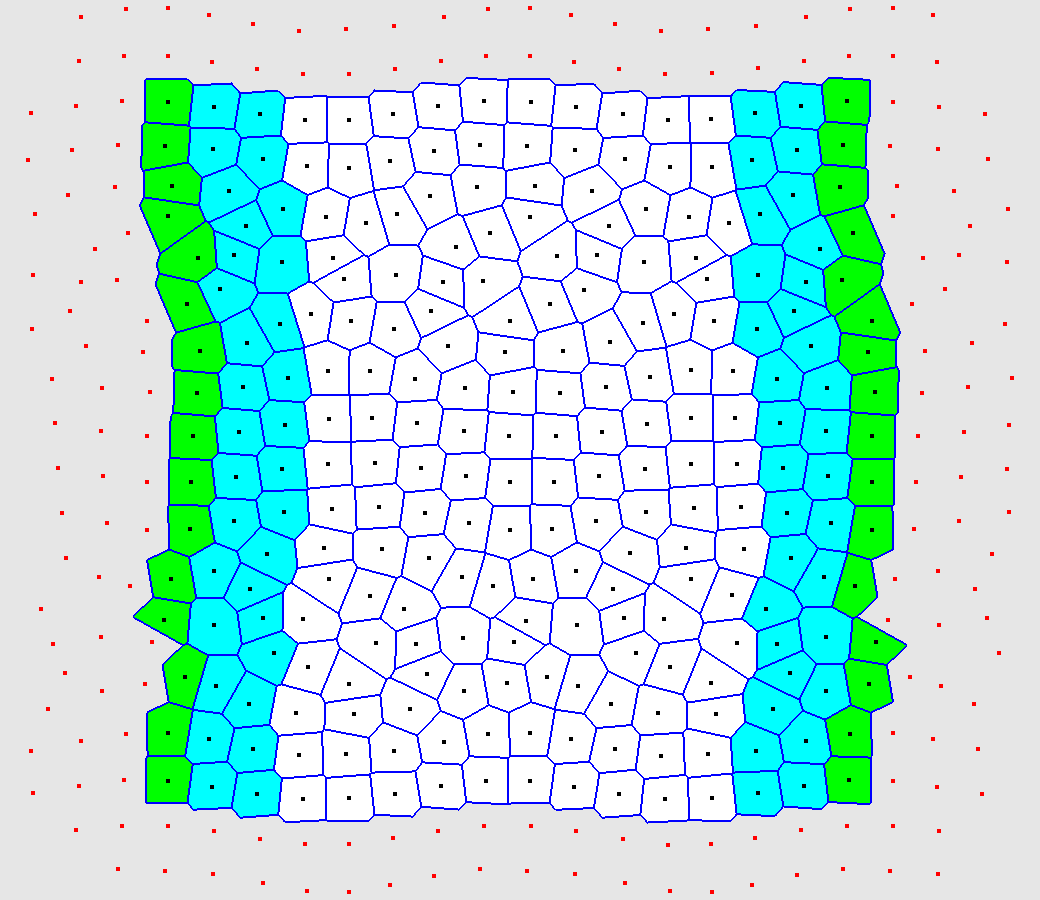}
\caption{ The next timestep's boundary cells are identified in the x dimension.  In this example, we are using periodic boundary conditions.  The green cells are the interior cells which share a face with the boundary.  In addition to these cells, we identify their neighbors and next-nearest neighbors (blue).  All of these cells are then copied and moved to become "ghost cells".
\label{fig:bcs} }
\end{figure}

We consider a single mesh generating point X and all of its ``friends of friends''.  First we find the nearest neighbor to X (which is guaranteed to share a face with X).  Call this neighbor Y.  Next we take the rest of potential neighbors and order the list by clockwise angle with respect to the segment XY.  What follows is an elimination process; at the end of this process, the elements in the list will be exactly those which share a face with X.  At each step in the process, we consider three consecutive elements of the list; call them A,B,C (see Fig. \ref{fig:tess1}).  We denote the element before A ``P'', and the element after C ``Q''.  It is determined whether or not to keep point B in the list, by checking whether C lies within the circle generated by X, A, and B.  If it does, then we remove B from the list, and take one step backward (checking whether C lies within XPA).  If C does not lie within the original circle, we keep B and move forward to check triangle BCQ).  More concisely, if ${\bigcirc_{XAB}}$ contains C, remove B from the list and take one step back: ${A \rightarrow P , B \rightarrow A}$.  Otherwise take one step forward: ${C \rightarrow Q, B \rightarrow C, A \rightarrow B}$.  Fig. \ref{fig:tess2} demonstrates an example of this full procedure.

Note that this algorithm is not sensitive to the presence of degenerate sets of points (that is, sets of four points which all lie on the same circle).  For practical purposes, it will not matter whether our code chooses to accept or reject a point in this configuration, because a degeneracy in the Delaunay tessellation corresponds to a face of zero area in the Voronoi tessellation.  If the face has zero area, then there will be zero flux through it, and hence it will have no influence on the resulting physics.

Once this operation has been performed for all members of the list (i.e., point B is now point Y, the nearest neighbor) all remaining list members are neighbors of point X.  All that remains is to calculate the areas and centers of mass of the faces, which is straightforward given the vertices of the polygon generated by this list.  These vertices are the centers of the circles generated by consecutive triples in the list. 

The details of the tessellation algorithm do not completely extend to three dimensions, but the idea of using friends-of-friends as a list of candidates still works, so in the worst case scenario we could use a brute force guess-and-check algorithm when D=3.  One might ask whether there is a major efficiency advantage to this tessellation procedure over more conventional ones such as direct insertion.  While it is not clear which method should be the fastest (given an optimized implementation), it is assumed they are comparably efficient.   Moreover, while the tessellation procedure takes up a non-negligible percentage of the code's overall runtime, it does not take up a $majority$ of the runtime, and as such there is no major incentive to optimize its efficiency.  The main advantage to the method described here is that the algorithm is very simple and does not require making a lot of exceptions.  Additionally, this algorithm is expected to be very easy to parallelize, because the tessellation is performed locally.  In the most simple prescription, we could make the code parallel via a simple domain decomposition, where different processes only share boundary data.  The main disadvantage to the tessellation procedure is that we must have an approximate tessellation to begin with, so that we can use ``friends of friends'' for our initial pool of neighbors.  In principle, this simply amounts to saving the tessellation from the previous timestep, but in practice this makes the implementation of boundary conditions a bit more complicated.  Not only do we need to create an appropriate set of ``ghost cells'' outside the boundary, but we need to generate ``ghost faces'' as well (the tessellation procedure won't automatically do this for us, because it relies on having an approximate tessellation from a previous timestep).  This also adds some complication to the parallelization of the code, for the same reasons. 

One might wonder why we worry so much about the positions of ghost cells.  For periodic boundary conditions, we don't really need ghost cells at all, since we could in principle associate left-most neighbors with right-most neighbors and so on, so that no boundary need be created.  For reflecting boundaries, one might hope we could have a fixed set of ghost cells lining our reflecting wall, so that we don't need to generate a new set each timestep.  Unfortunately, we only have control over the positions of mesh generators; we don't directly control shape of the cells.  Thus, if we want a flat wall, we need to have the mesh generators reflected across the wall.  The only reason we include ghost cells for periodic boundary conditions is for the sake of overall simplicity in the code (fewer parts of the code depend on the choice of boundary conditions).  In this case, ghost cells are translated to the opposite end of the domain with respect to their ``real'' counterparts.

The boundary conditions are set in each dimension sequentially.  The first step involves flagging cells according to whether they are inside or outside the boundary.  If we are using periodic boundary conditions and a cell moves off of the computational domain, it is set to be a ghost cell and its corresponding ghost cell is set to be a real cell.  All cells are flagged to be in one of three categories:  Inside the domain, outside the domain, or a ``border cell'', meaning it is inside the domain and neighbors a cell outside the domain.

The next step involves generating a new list of ghost cells, first by making copies of all border cells.  Additional ghost cells get added to this list, by using all neighbors of ghost cells which are inside the computational domain.  This procedure is repeated until the desired number of ghost cells is achieved (see Fig. \ref{fig:bcs}).  For our second-order methods, we need two layers of ghost cells.

Next these copies are moved according to the boundary conditions, e.g. if the boundaries are reflecting, their positions are reflected across the reflecting wall.

Finally, neighbors are assigned to the copies by using the neighbor data from the original tessellation.  These associations are of two kinds: associations between two copied cells, and associations between a copied cell and a border cell.  Both kinds of associations can be extracted from the original tessellation.

After this step, we discard all the old ghost cells and replace them with the new ghost cells.  When this is done, the tessellation algorithm is performed as previously described.  This implementation of boundary conditions is essentially the same as the method used in AREPO, though we require a bit more care because we need to retain the tessellation information in the boundary cells.  As a final note on this formulation, very little of the code depends on the number of dimensions; only the tessellation algorithm itself is significantly affected by D.

\subsection{Mesh Regularity}

For most problems, evolving the mesh according to the above prescription can generically lead to cells which are long and skinny, and whose mesh generating points are very close to their edges.  These cells will evolve in an unstable manner, because their faces can move very quickly even while their mesh generators are moving slowly, and they can also have a very short sound-crossing time.  It is therefore desirable to steer cells in such a way that they tend to become more regular.  \cite{arepo} found an effective prescription for this, which is to give the mesh generators an additional component to their velocity, pointed in the direction of the center of mass.  We repeat this prescription here:

\begin{equation}
\vec w'_i = \vec w_i + \chi \left\{ \begin{array}
                  {l@{\quad:\quad}l}
                 0 & d_i / (\eta R_i) < 0.9  \\  
                 c_i {\vec s_i - \vec r_i \over d_i} {d_i - 0.9 \eta R_i \over 0.2 \eta R_i} & 0.9 < d_i / (\eta R_i) < 1.1  \\
                 c_i {\vec s_i - \vec r_i \over d_i} & 1.1 < d_i / (\eta R_i)
                  \end{array} \right.
\end{equation}

${R_i = \sqrt{dV_i/\pi}}$ is the effective radius of cell i, ${d_i}$ is the distance between the cell's mesh generating point ${\vec r_i}$ and its center of mass ${\vec s_i}$.  ${c_i}$ is the local sound speed.  $\eta$ and $\chi$ are arbitrary parameters for this prescription, which are typically set to ${\eta = 0.25}$ and ${\chi = 1.0}$ (the same values typically used by AREPO).  Note that we we do not implement a relativistic velocity addition formula, as ${\vec w}$ need not be interpreted as a physical velocity.

\section{Test Problems}
\label{sec:test}

\begin{figure*}
\epsscale{1.0}
\plotone{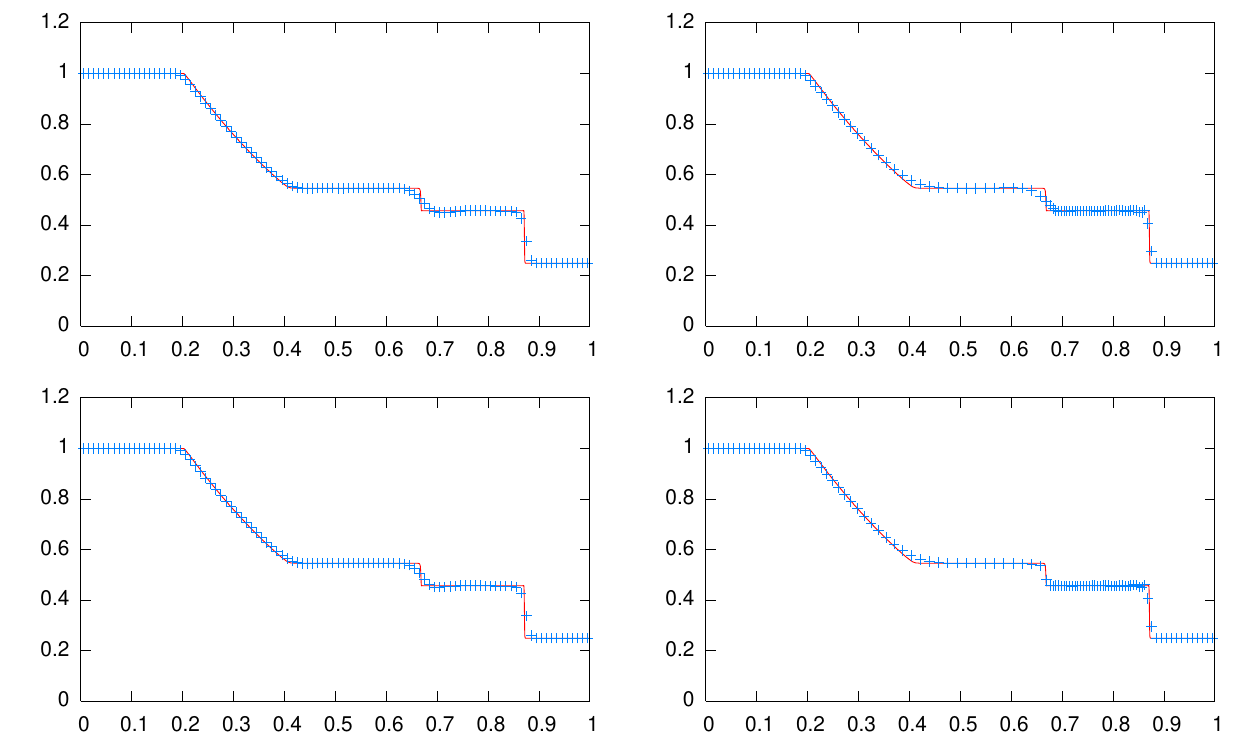}
\caption{ Rest mass density in the nonrelativistic shock tube at ${t = 3.0}$ with 100 cells.  Results for a static mesh on the left, and moving mesh on the right.  Top plots use the HLL Riemann solver, bottom plots use HLLC.  The solid line represents the exact solution.  The contact discontinuity is much better preserved when employing HLLC on the moving mesh.
\label{fig:shock1a} }
\end{figure*}

As this is a new method for solving relativistic hydrodynamics, we present a large number of test problems in one and two dimensions.  In addition to the relativistic tests, we have included some nonrelativistic ones, to compare our code with AREPO.

\subsection{One-Dimensional Tests}
\begin{deluxetable}{ccccc}
\tablecaption{One-Dimensional Tests
\label{tab:1d}}
\tablewidth{0pt}
\tablehead{
\colhead{Test}    &   \colhead{Variable}
}
\startdata
Nonrelativistic &  & ${x<.5}$  & ${x>.5}$ \\ 
Shock Tube & $\rho$ & 1  & .25 \\ 
N = 100 & v    & 0  & 0 \\ 
${\Gamma = 1.4}$ & P  & 1  & .1795\\
t = 3.0 & & &\\
\tableline
Nonrelativistic  &   & ${x<.1}$  &  ${.1<x<.9}$ & ${x>.9}$ \\
Interacting Shocks & $\rho$ & 1  & 1 & 1\\ 
N = 400 & v    & 0 & 0 & 0 \\ 
${\Gamma = 1.4}$ & P  & 1000  & .01 & 100 \\
t = .038 & & &\\
\tableline
Easy Relativistic &   &   ${x<.5}$ & ${x>.5}$ \\
Shock Tube & $\rho$ & 1  & 1 \\ 
 & vx    & 0 & 0 \\ 
N = 400 & vy    & 0 & .99 \\ 
${\Gamma = 5/3}$ & P  & 1000 & .01\\ 
t = 0.4 & & &\\
\tableline
Hard Relativistic &  &   ${x<.5}$ & ${x>.5}$ \\
Shock Tube & $\rho$ & 1  & 1 \\ 
 & vx    & 0 & 0 \\ 
N = 100 & vy    & .9 & .9 \\ 
${\Gamma = 5/3}$ & P  & 1000 & .01\\
t = 0.6 & & &
\enddata
\end{deluxetable}

All 1D tests involving piecewise constant states are summarized in table \ref{tab:1d}.  Our first two tests are identical to tests performed by \cite{arepo}.  The first is a simple shock tube, a test which has also been performed by \cite{shock1}.  To demonstrate the importance of the HLLC Riemann solver in our scheme, we perform four tests, varying whether the mesh is static or moving, and varying whether we use HLL or HLLC.  Results are plotted in Fig. \ref{fig:shock1a}.  When the cells were moved and HLLC was employed, the contact discontinuity was very well approximated.   In the other three cases, there was far more diffusion of the contact discontinuity.  This demonstrates that the accuracy in this scheme comes from the combination of using a moving mesh and employing a multi-state Riemann solver.

\begin{figure}
\epsscale{1.0}
\plotone{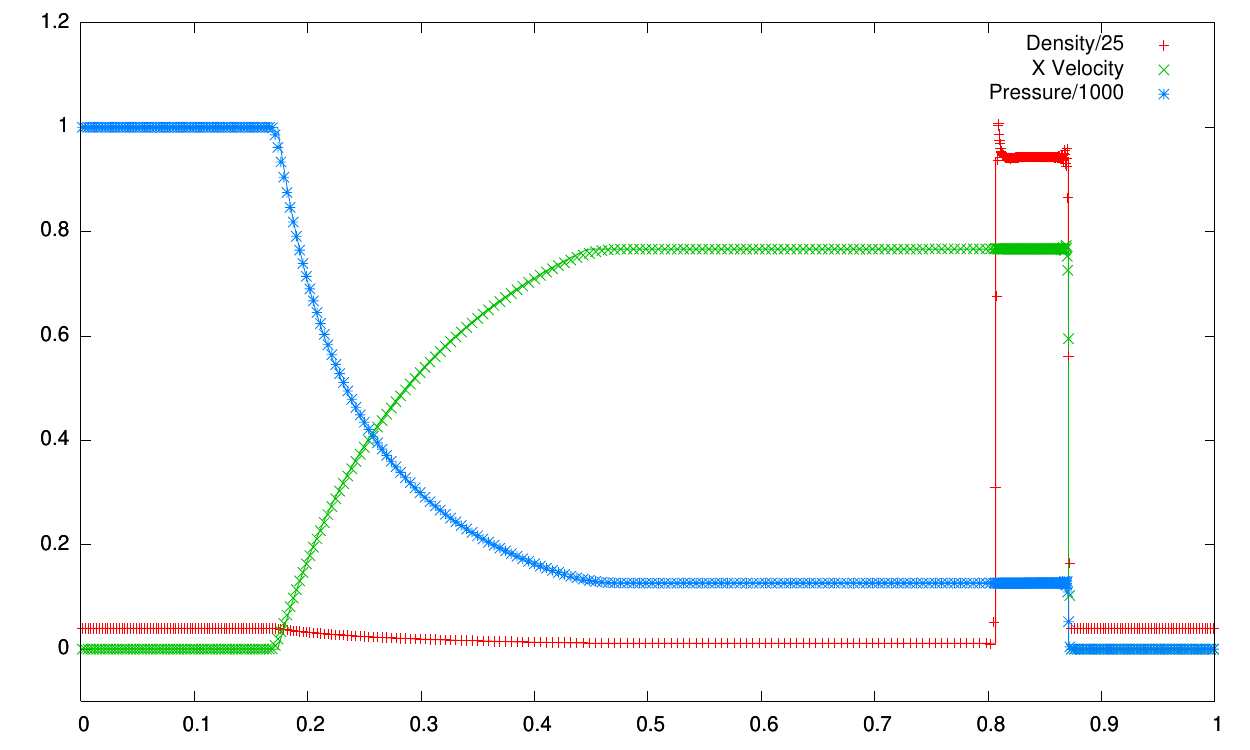}
\caption{ Easy relativistic shock tube at ${t=0.4}$ with 400 cells using a moving mesh.  Solid lines represent the exact solution.
\label{fig:shock2} }
\end{figure}

\begin{figure*}
\epsscale{1.0}
\plotone{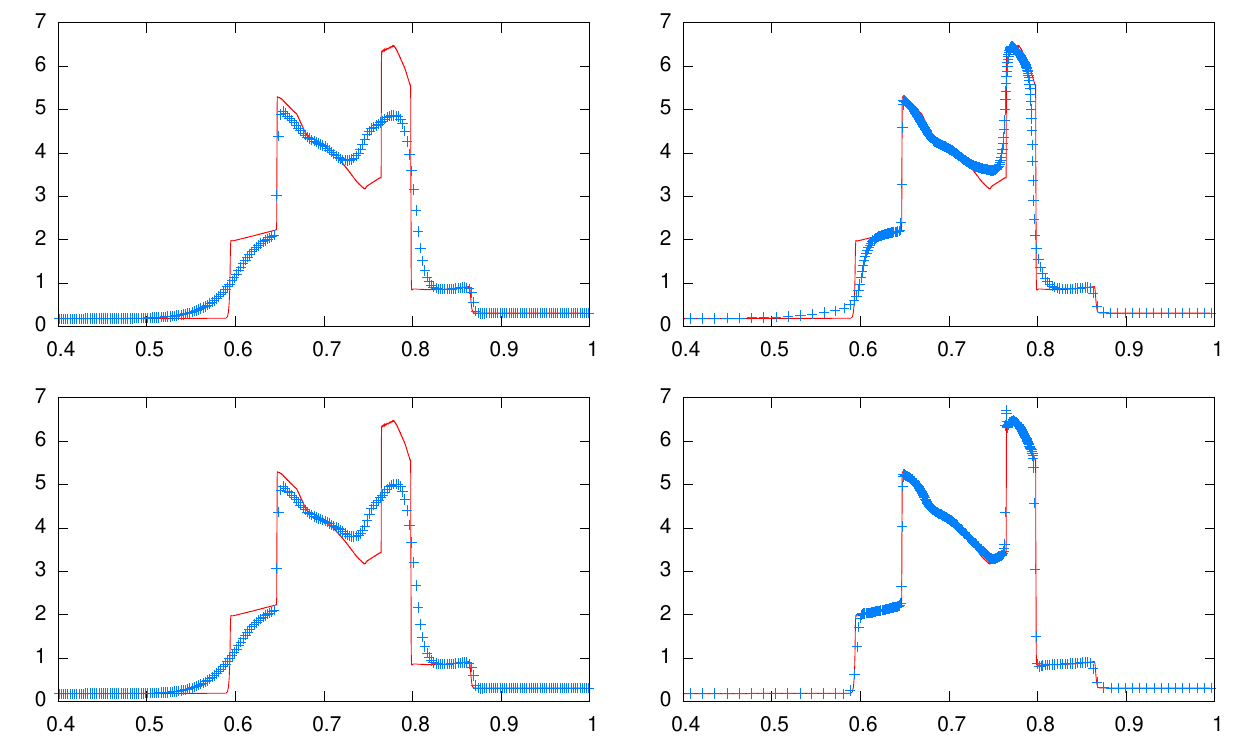}
\caption{ Rest mass density in the double blastwave test at ${t = .038}$ with 400 cells.  Results for a static mesh on the left, and moving mesh on the right.  Top plots use the HLL Riemann solver, bottom plots use HLLC.  The solid line is meant to represent the exact solution; it was calculated numerically at much higher resolution.  Moving the mesh improves accuracy, and using HLLC on this mesh brings a vast improvement in resolving both shocks and contact discontinuities.
\label{fig:dblast} }
\end{figure*}

The second nonrelativistic test we perform involves the interaction of multiple shocks \citep{dblast}.  Fig. \ref{fig:dblast} shows the multiple-shock problem at ${t=.038}$.  We compare results with a fixed and moving mesh, and using the HLL and HLLC Riemann solvers.  Again, we see that the combination of using the HLLC solver and allowing the cells to move leads to a very high accuracy in the solution.

\begin{deluxetable}{cccc}
\tablecaption{$L_1$ errors of the density for the easy relativistic shock tube at $t=0.4$.  TESS is compared with RAM \tablenotemark{*}.
\label{tab:easy}}
\tablewidth{0pt}
\tablehead{
\colhead{Code}    &   \colhead{$N$}  &  
\colhead{$L_1$ Error} & \colhead{Convergence Rate}
}
\startdata
TESS & 100  & 4.23e-1 &      \\ 
       & 200  & 2.57e-1 & 0.82 \\ 
       & 400  & 1.36e-1 & 0.82 \\ 
       & 800  & 7.45e-2 & 0.98 \\
       & 1600 & 3.54e-2 & 0.91 \\ 
       & 3200 & 2.48e-2 & 0.95 \\ 
\tableline
RAM  & 100  & 8.48e-1 &      \\ 
       & 200  & 4.25e-1 & 1.0  \\ 
       & 400  & 2.41e-1 & 0.82  \\ 
       & 800  & 1.27e-1 & 0.92 \\ 
       & 1600 & 6.43e-2 & 0.99  \\ 
       & 3200 & 3.34e-2 & 0.95
\enddata
\tablenotetext{*}{RAM used piecewise parabolic reconstruction and a modified Marquina flux}
\end{deluxetable}

\begin{figure}
\epsscale{1.0}
\plotone{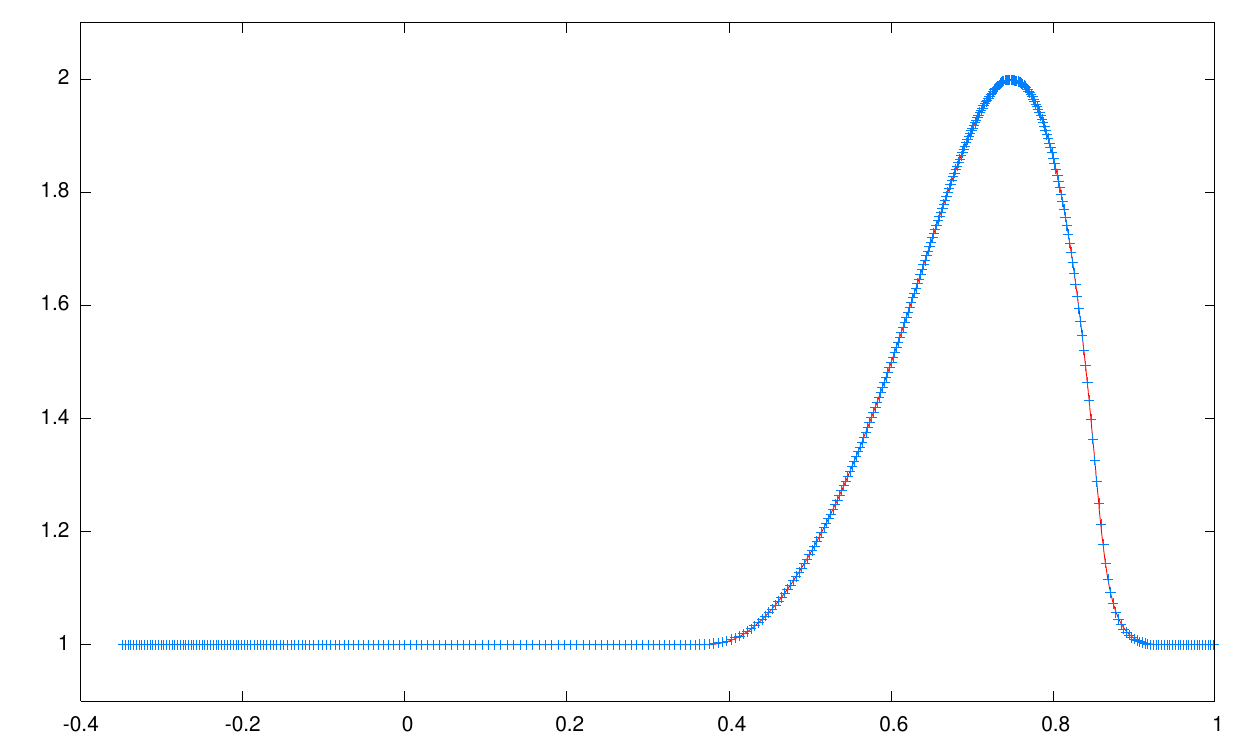}
\caption{ Isentropic pulse at ${t=0.8}$ with 400 cells; a smooth problem to test second-order convergence.  The solid line was calculated at much higher resolution.
\label{fig:isentropic} }
\end{figure}

\begin{figure*}
\epsscale{1.0}
\plotone{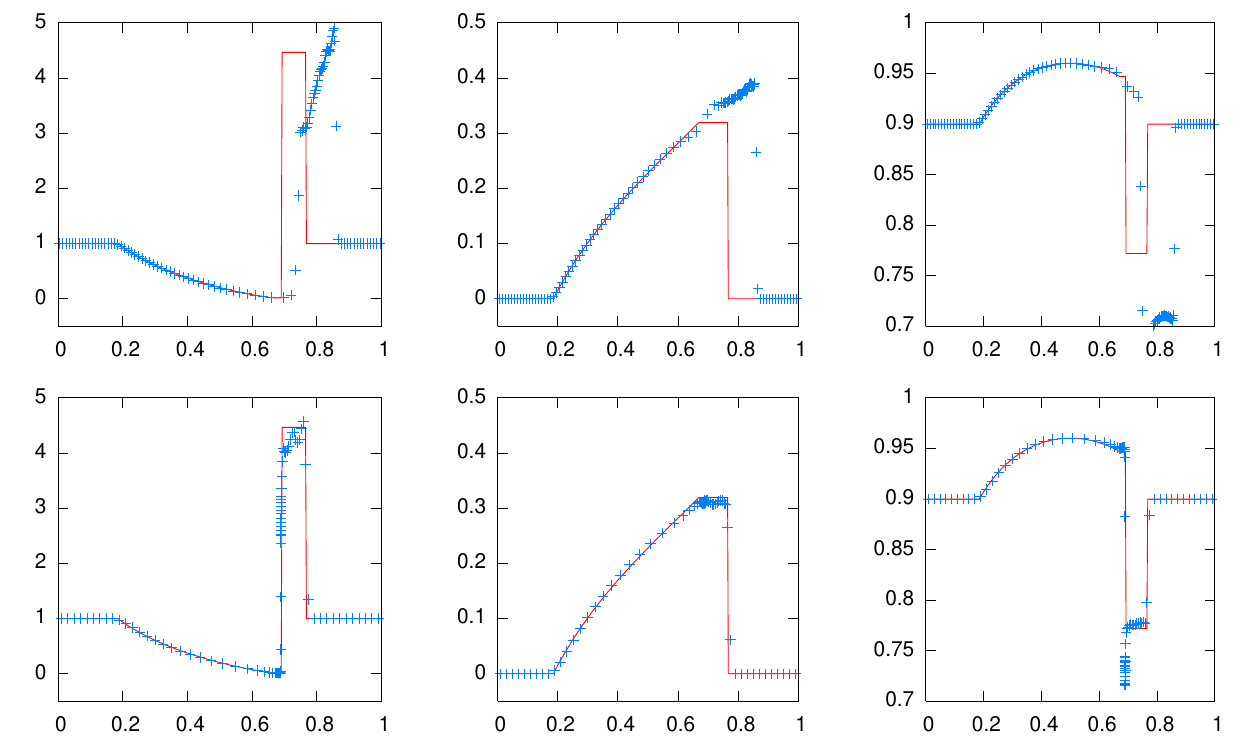}
\caption{ Hard relativistic shock tube at ${t=0.6}$ (solid line is the exact solution).  Both tests were run with a moving mesh using 100 zones, but the lower test concentrated 50 of the zones within a small region near the initial discontinuity, momentarily giving an effective resolution of roughly 10,000.
\label{fig:shock3} }
\end{figure*}

For relativistic one-dimensional tests, we compare our code to the relativistic adaptive mesh refinement code RAM \citep{ram} which was tested for convergence against a variety of test problems.  The first two are Riemann problems with transverse velocity, the so-called ``Easy'' and ``Hard'' shock tube tests.  The easy shock tube test is plotted in Fig. \ref{fig:shock2} at time ${t=0.4}$.  For the easy shock tube, we find nearly first-order convergence, and smaller errors than RAM (see table \ref{tab:easy} for convergence rates).  For the hard shock tube, we find that like RAM, we need very high resolution to capture the solution accurately.  However, because we can initially place the cells anywhere we want, we can resolve the initial discontinuity very well and capture the solution with as few as 100 conputational zones (see Fig. \ref{fig:shock3}).  50 of the zones were concentrated uniformly within a region ${\Delta x = .005}$ of the discontinuity and the remaining 50 cells were distributed uniformly through the rest of the domain.  Using a uniform initial mesh, TESS showed first order convergence for this problem (table \ref{tab:hard}).

\begin{deluxetable}{cccc}
\tablecaption{$L_1$ errors of the density for the hard relativistic shock tube.  TESS is compared with RAM\tablenotemark{*}
\label{tab:hard}}
\tablewidth{0pt}
\tablehead{
\colhead{Code}       &
\colhead{$N$}  &  
\colhead{$L_1$ Error} & \colhead{Convergence Rate}
}
\startdata
TESS    &    400  &  7.12e-1  &       \\
           &    800  &  4.54e-1  &  0.64 \\
           &   1600  &  2.26e-1  &  1.0 \\
           &   3200  &  1.10e-1  &  1.0 \\
\tableline
RAM    &    400  &  5.21e-1  &       \\
           &    800  &  3.63e-1  &  0.52 \\
           &   1600  &  2.33e-1  &  0.64 \\
           &   3200  &  1.26e-1  &  0.89
\enddata
\tablenotetext{*}{RAM used piecewise parabolic reconstruction and a modified Marquina flux}
\end{deluxetable}

The last test we perform in one dimension is to demonstrate the convergence rate on a smooth problem.  We set up an isentropic pulse identical to the one used by \cite{ram}:

\begin{equation}
\rho = \rho_{ref} ( 1 + \alpha f(x) )
\end{equation}
\begin{equation}
f(x) =  \left\{ \begin{array}
                  {l@{\quad:\quad}l}
                 ((x/L)^2 - 1)^4 & |x| < L  \\  
                 0 & otherwise
                  \end{array} \right.
\end{equation}
\begin{equation}
P = K \rho^\Gamma
\end{equation}

${\rho_{ref}}$, ${K}$,${L}$, and ${\alpha}$ are constants.  In our case, ${\rho_{ref} = 1.0}$ ${K = 100}$, ${L = 0.3}$, ${\alpha = 1}$.  ${\Gamma}$ is the adiabatic index, chosen to be ${5/3}$.  To determine the velocity, we use

\begin{eqnarray}
  J_- = \frac{1}{2} \ln (\frac{1+v}{1-v}) - 
  \frac{1}{\sqrt{\Gamma-1}}
  \ln (\frac{\sqrt{\Gamma-1}+c_s}{\sqrt{\Gamma-1}-c_s} ) \\
  \nonumber \\
  = constant \nonumber
  \label{eqn:rmninv}
\end{eqnarray} 

where ${c_s}$ is the sound speed.

\begin{deluxetable}{cccc}
\tablecaption{$L_1$ errors of the density for the isentropic pulse.  TESS is compared with two versions of RAM.
\label{tab:isentropic}}
\tablewidth{0pt}
\tablehead{
\colhead{Code}       &
\colhead{$N$}  &  
\colhead{$L_1$ Error} & \colhead{Convergence Rate}
}
\startdata
TESS  &   80 &    4.88e-3   &    \\
           &   160 &   1.78e-3   &     1.8  \\
 	  &   320 &   4.84e-4   &    1.9  \\
 	  &   640 &   1.20e-4   &    2.0  \\
	  &  1280 &   2.93e-5   &    2.1  \\
	  &  2560 &   6.97e-6   &    2.1  \\
	  &  5120 &   1.47e-6   &    2.1  \\
\tableline
RAM  &   80 &    1.12e-2   &    \\
(U-PLM-RK4) &   160 &   3.56e-3   &     1.7  \\
 	  &   320 &   1.02e-3   &    1.8  \\
 	  &   640 &   2.60e-4   &    2.0  \\
	  &  1280 &   6.49e-5   &    2.0  \\
	  &  2560 &   1.62e-5   &    2.0  \\
	  &  5120 &   4.04e-6   &    2.0  \\
\tableline
RAM  &   80 &   1.10e-2  &    \\
(U-PPM-RK4) &  160 &   2.56e-3  &     2.1  \\
 	  &  320 &   5.74e-4  &     2.2  \\
 	  &  640 &   1.34e-4   &    2.1  \\
	  &  1280 &  3.10e-5  &     2.1  \\
	  &  2560 &  7.33e-6  &     2.1  \\
	 &  5120 &   1.82e-6   &    2.1
\enddata
\end{deluxetable}

Fig. \ref{fig:isentropic} shows the pulse at ${t=0.8}$ on a mesh with 400 cells.  The L1 error and convergence rates are shown in table \ref{tab:isentropic}.  In order to make a reasonable comparison between the codes, we chose two different methods employed by RAM which are second order.  Of course, if we had chosen to compare with the WENO solvers employed by RAM, it would be no contest, as RAM can get up to fifth order convergence for smooth flows.  For this problem, we not only find smaller errors, but also slightly faster convergence than RAM (RAM employed piecewise linear reconstruction of the primitive variables, and a fourth-order Runge-Kutta time integration, leading to an overall second-order scheme).  In fact, convergence for TESS was slightly better than second order.  This could be due to the fact that the method becomes ``more Lagrangian'' as the resolution increases; face velocities approach the velocities of contact waves in this limit.

\subsection{Convergence in Multiple Dimensions}

On the surface, it is certainly not obvious that this second-order convergence extends to multiple dimensions; it would be quite difficult to prove such a thing mathematically.  A convergence test in 2D is essential if we want to establish confidence in our numerical scheme.  A straightforward test which satisfies this is to propagate an isentropic wave diagonally across the mesh.  The mesh will initially be a square grid, but because the flow is nonuniform, the mesh will move in a nontrivial way and we will be able to check whether second-order convergence continues to hold during such distortions.  The result we present will employ a nonrelativistic wave in a periodic box ${(x,y) \in [0,1] \times [0,1]}$, with density and pressure given by the same formulas as in the 1D case, but now with

\begin{equation}
f(x,y) = sin^2( \pi (x+y) ).
\end{equation} 

For this test, we use ${K=.01}$, but all other constants are the same as in the previous example.  The nonrelativistic limit of equation (\ref{eqn:rmninv}) is simply

\begin{equation}
v = {2 \over \Gamma-1} ( c_s - c_s^{ref} ).
\label{eqn:isenv}
\end{equation}

The direction of this velocity is of course diagonal to the mesh,

\begin{equation}
\vec v = v( \hat x + \hat y )/\sqrt{2}.
\end{equation}

\begin{figure}
\epsscale{1.0}
\plotone{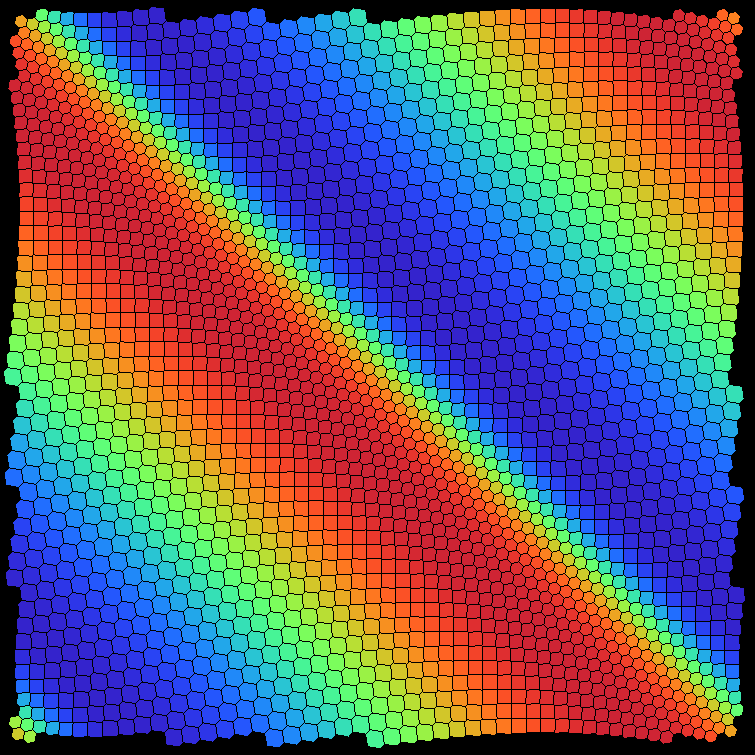}
\caption{ The 2D isentropic wave at t = 1.0 on a ${50 \times 50}$ mesh.  This test demonstrates second-order convergence when the mesh is allowed to move and dynamically change its geometry (initially we start with a uniform grid).
\label{fig:isen2d} }
\end{figure}

\begin{figure}
\epsscale{1.0}
\plotone{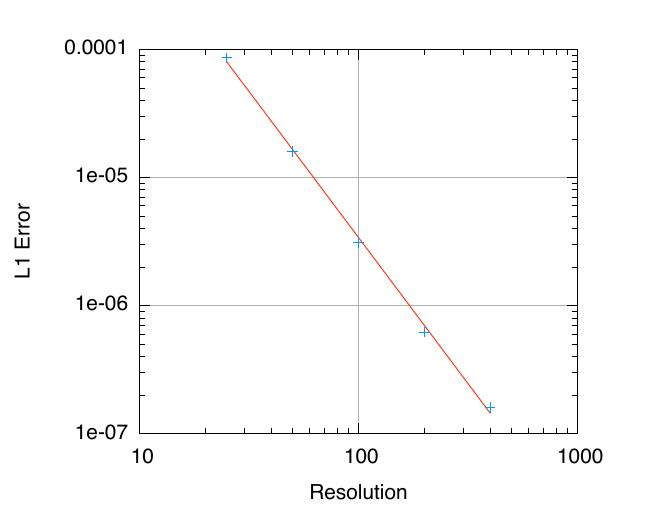}
\caption{ L1 error in the velocity for the 2D isentropic wave at t = 1.0 at various resolutions.  The slope of the solid line is 2.3.
\label{fig:l12d} }
\end{figure}

Condition (\ref{eqn:isenv}) continues to hold as the wave evolves, so we can use this as a way of measuring L1 error for this problem.  In Fig. \ref{fig:isen2d} we see the pulse at ${t=1.0}$ on a ${50 \times 50}$ mesh, and the cells have clearly changed their shape and size in a nontrivial way.  L1 error for this problem is plotted in Fig. \ref{fig:l12d}.  The convergence rate for this problem was calculated to be 2.3, again slightly better than second order.

\subsection{Tests in Spherical and Cylindrical Coordinates}

TESS is written in a modular way so that it is straightforward to change the set of conserved quantities, and to add source terms.  As an example, we have implemented the equations in alternative coordinate systems for the two special cases of spherical symmetry and axisymmetry.  Our extension to these coordinate systems takes the most na\"ive approach: we simply express the equations in an alternate coordinate system, and solve them like any other hyperbolic system in conservation-law form.  We do not wish to endure the complication of curvilinear voronoi cells, which is why we specialize to the cases of spherical symmetry (1-D) and axisymmetry (2-D).  For example, in cylindrical coordinates in axisymmetry the nonrelativistic conservation laws take the following form:

\begin{equation}
U = \left( \begin{array}{c} r \rho \\ r \rho v_r \\ r \rho v_z \\ r^2 \rho v_\phi \\ r( \rho v^2/2 + \rho e ) \end{array} \right)  
\end{equation}
\begin{equation}
\vec F \cdot \hat n = \left( \begin{array}{c} r \rho \vec v \cdot \hat n \\ r ( \rho v_r \vec v \cdot \hat n + P \hat n_r ) \\ r ( \rho v_z \vec v \cdot \hat n + P \hat n_z ) \\ r^2 \rho v_\phi \vec v \cdot \hat n  \\ r ( \rho v^2/2 + \rho e + P ) \vec v \cdot \hat n \end{array} \right) 
\end{equation}
\begin{equation}
S = \left( \begin{array}{c} 0 \\ \rho v_\phi^2 + P  \\ 0 \\ 0 \\ 0 \end{array} \right)
\end{equation}

where $r$ is the distance from the z-axis (note that ${\hat n}$ lives in the r-z plane).  The two new developments here are the presence of the radial coordinate $r$ in the equations and the non-zero source term.  We evaluate both of these at the center of mass of a cell or face, depending on the context.  The description in spherical coordinates and the extensions to relativistic hydrodynamics are completely analogous.


We use this opportunity to test the method's extension to these coordinate systems, as well as the ability for TESS to preserve symmetries using an unstructured mesh.  We set up very simple shock-tube like initial conditions:

\begin{equation}
\rho =  \left\{ \begin{array}
                  {l@{\quad:\quad}l}
                 1.0 & r < .25  \\  
                 0.1 & r > .25
                  \end{array} \right.
\end{equation}
\begin{equation}
P = \rho
\end{equation}
\begin{equation}
v  = 0
\end{equation}

\begin{figure}
\epsscale{1.0}
\plotone{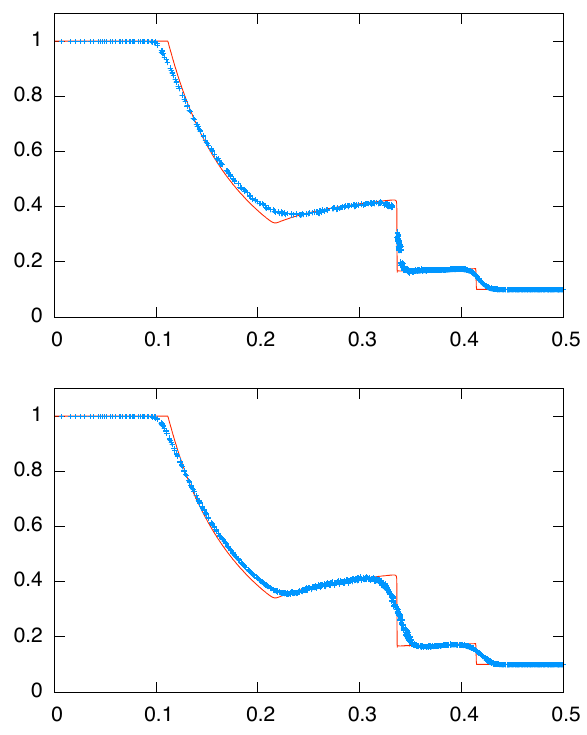}
\caption{ Density profile for the relativistic cylindrical explosion using a 50 $\times$ 50 cartesian mesh at ${t=0.2}$.  The plot above uses a fixed mesh, while the plot below uses a moving mesh.  The solid line was calculated in one dimension (cylindrical coordinates) at high resolution.  The cells plotted lie along a single line; blue pluses represent cells that lie along the x-axis and purple crosses represent cells that lie along the diagonal ${x=y}$.
\label{fig:cyl} }
\end{figure}

\begin{figure}
\epsscale{1.0}
\plotone{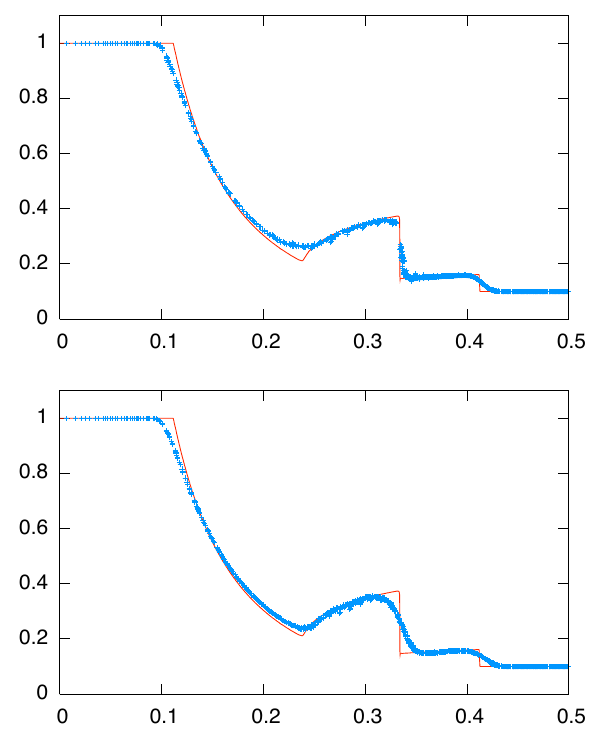}
\caption{ Density profile for the relativistic spherical explosion using a 50 $\times$ 50 mesh in cylindrical coordinates at ${t=0.2}$.  Conventions are the same as in Figure \ref{fig:cyl}.
\label{fig:sph} }
\end{figure}

We perform a relativistic cylindrical and spherical explosion.  For the cylindrical explosion, we perform the calculation at high resolution in 1-d using cylindrical coordinates, and in low resolution in 2D using cartesian coordinates.  The resolution is as low as 50 $\times$ 50.  In Fig. \ref{fig:cyl} we show the results for a moving and fixed mesh.  We see that moving the cells continues to improve resolution of the contact discontinuity, and in this case we also see some improvement in symmetry -- the values for the density profile are not as scattered in the moving-mesh case, they tend to lie along a single curve.  For the spherical explosion (Fig. \ref{fig:sph}), we see a similar story; the contact discontinuity is well preserved, and the spherical symmetry is captured somewhat better with the moving code.

\subsection{Test Problems in Magnetohydrodyamics}

To further illustrate that TESS in principle extends to an arbitrary set of conserved quantities, we demonstrate a few test problems in magnetohydrodynamics.  From one point of view, MHD (whether relativistic or nonrelativistic) is just another hyperbolic set of equations in conservation-law form, where electromagnetic energy, momentum, stress, and pressure are added to the fluid equations, and the magnetic induction also satisfies a conservation law (Faraday's Law), which in the the case of a perfect conductor (${\vec E = -\vec v \times \vec B}$) has the form

\begin{equation}
\partial_t \vec B + \partial_j ( v^j \vec B - B^j \vec v ) = 0
\end{equation}

The main ingredient which distinguishes MHD from other conservation laws is that the equations also satisfy an elliptic constraint, ${\nabla \cdot B = 0}$, which must be satisfied every timestep.  This constraint is analytically guaranteed to vanish, but depending on your numerical scheme, the constraint can generically grow exponentially when the equations are solved in discrete form.  Over the years, many techniques have been suggested to address this issue, the most popular methods being variants of constrained transport \citep{toth}, where steps are taken to ensure that all fluxes added to the B fields have zero divergence, so that the time derivative of div B is guaranteed to remain zero through the entire evolution (in other words, only a solenoidal B field is ever added during the time evolution).  This method would be quite difficult to extend to an unstructured mesh, so it seems that we must choose some other way of preventing these constraint-violating solutions from growing.

We employ a hyperbolic divergence-cleaning scheme \citep{telegraph} which adds an extra conserved quantity, $\psi$, and modifies the equations so that the evolution equations for the magnetic field become

\begin{eqnarray}
\partial_t  \vec B + \nabla_j ( v_j \vec B - B_j \vec v ) + \vec \nabla \psi = 0 \\
\partial_t \psi + c^2_h \nabla \cdot B = - {c^2_h \over c^2_p} \psi
\end{eqnarray}

where ${c^2_h}$ and ${c^2_p}$ are freely specifiable constants which alter the behavior of the divergence-cleaning terms.  If ${\nabla \cdot B = 0}$ and ${\psi=0}$ the equations revert to the usual MHD equations, but this introduction of $\psi$ has the effect of altering the time evolution of the constraint so that it satisfies a wave equation, specifically the telegraph equation.  This technique is generally not as desirable as constrained transport, partly because it doesn't guarantee zero divergence, but also because its success is strongly dependent on two freely specifiable constants, which have no obvious choice for what their values should be.  Because we don't have a constrained transport scheme (though we do speculate that one is possible), and also because this method is easy to implement, we currently use this divergence-cleaning form of the equations in our MHD code.

We now look at several MHD tests.  The first is a one-dimensional shock tube test in relativistic MHD, the relativistic version of the Brio-Wu shock tube \citep{bw88, vp93}.  The left and right states are as follows:

\begin{equation}
\rho_L = 1, \qquad \rho_R = .125 
\end{equation}
\begin{equation}
P_L = 1, \qquad P_R = .1
\end{equation}

The initial velocity is zero everywhere, and the magnetic field is given by the following:

\begin{equation}
B_x = 0.5, \qquad B_z = 0 
\end{equation}
\begin{equation}
B_{yL} = 1, \qquad B_{yR} = -1
\end{equation}

The adiabatic index ${\Gamma = 2}$.  The calculation was done with 400 zones.  The final state is shown at ${t = 0.4}$, in Fig. \ref{fig:briowu}.  It is clear that an advantage is gained in moving the cells, both in the sharpness of the contact discontinuity and in the shock front.  Additionally, the value of the density in the shock wave appears to be more accurate when using a moving mesh.  So, it would seem that there is great potential in using this Lagrangian approach for problems in magnetohydrodynamics.  However, we wish to avoid reading too much from this result, as this is a one-dimensional problem, and most of the challenges in magnetohydrodynamics only appear when solving multidimensional problems.

We consider now a test problem in nonrelativistic magnetohydrodynamics, a cylindrical explosion in a uniform magnetic field.  The calculation is done in a box of dimensions [0,1] x [0,1].  The initial conditions are as follows:

\begin{figure}
\epsscale{1.0}
\plotone{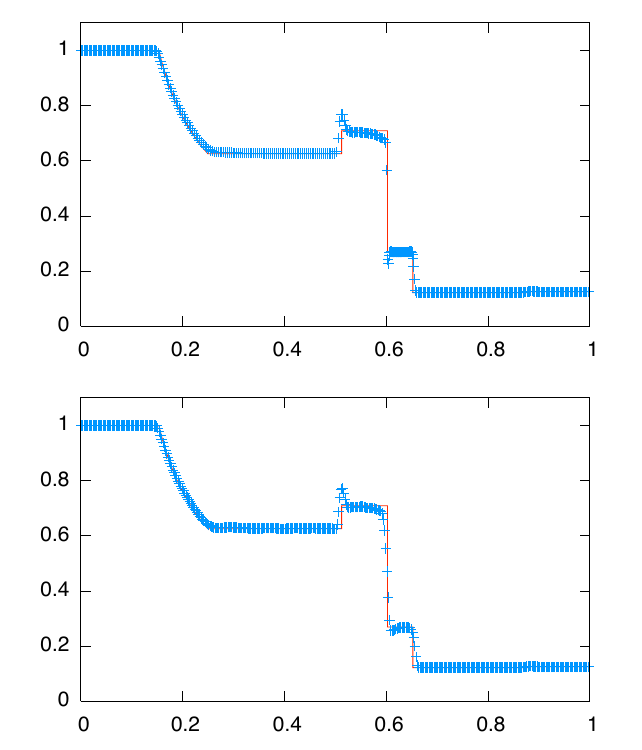}
\caption{ Density for the Brio-Wu relativistic MHD shock tube at t = 0.4 with 400 cells.  The upper plot uses a fixed mesh, and the lower plot uses a moving mesh.  The solid red line is the analytic solution.  Just as in the nonrelativistic hydrodynamical case, the contact discontinuity is captured much better on the moving mesh.
\label{fig:briowu} }
\end{figure}

\begin{equation}
\rho = 1.0
\end{equation}
\begin{equation}
B_x = 1.0
\end{equation}
\begin{equation}
P = \left\{ \begin{array}
                  {l@{\quad:\quad}l}
                 10.0 & r < 0.1  \\  
                 0.1   & r > 0.1
                  \end{array} \right.
\end{equation}

\begin{figure}
\epsscale{1.0}
\plotone{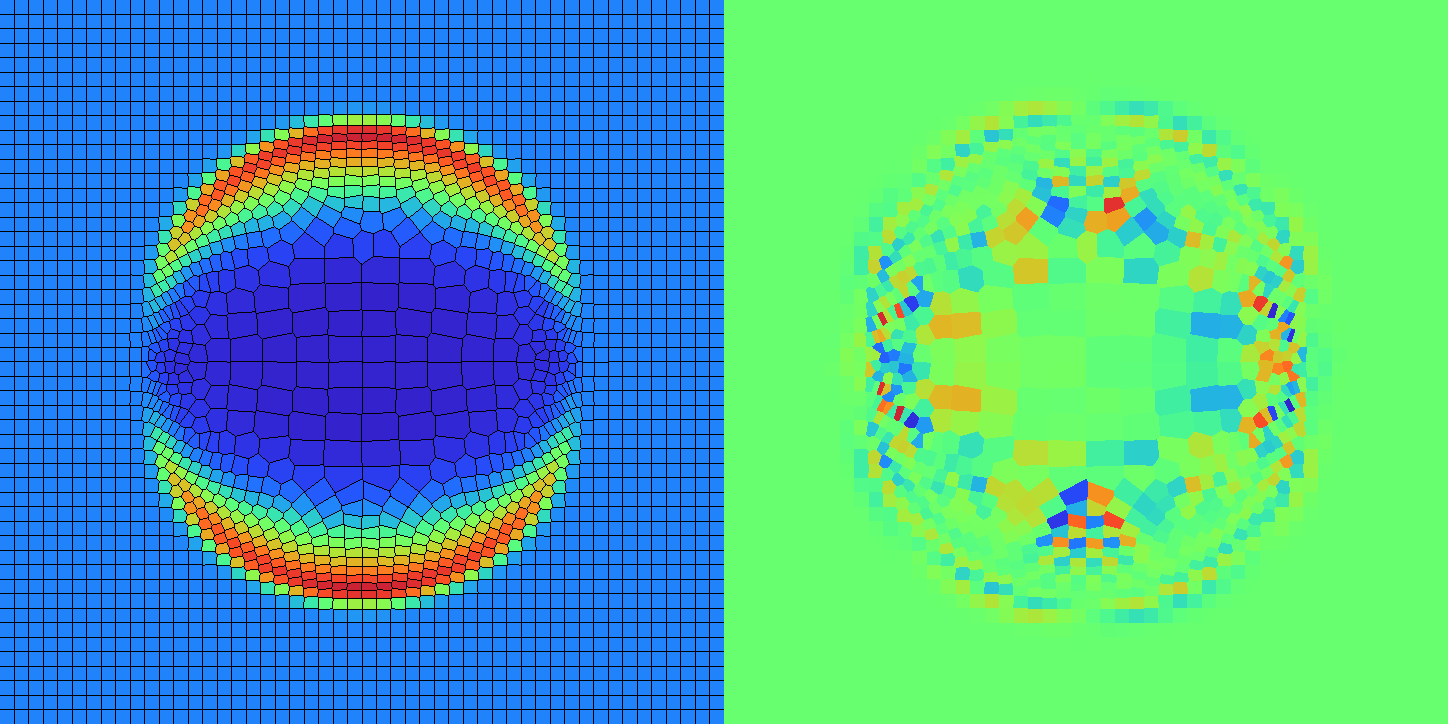}
\caption{ Nonrelativistic cylindrical explosion in a uniform magnetic field at t=0.1.  Magnetic pressure is plotted in the left panel, while normalized div B is plotted in the right panel.  Magnetic divergence grows to be unfortunately large; the normalized divergence is of order 10 percent.
\label{fig:mag} }
\end{figure}

The magnetic pressure at t=0.1 is plotted in Fig. \ref{fig:mag}.  Also plotted is the numerically calculated divergence of the magnetic field.  This divergence is normalized in the following manner:

\begin{equation}
(div B)_{norm} = {\nabla \cdot B \over |B|/R }
\end{equation}

Where R is the characteristic size of the Voronoi cell.  We see the code does a reasonable job of resolving the explosion at low resolution, however the magnetic divergence is unacceptably large (the normalized magnetic divergence grows to be as large as 0.14).  This large error in div B appears to be particularly problematic for this method.  Any time the Voronoi mesh changes topology, there can be extremely rapid growth in magnetic divergence due to the appearance of new faces in the tessellation.  This magnetic growth is so rapid that it shows up even in simple test problems like the one displayed.  In fact, the growth rate is such that conventional divergence-cleaning techniques for suppressing the growth of div B seem to be insufficient to resolve the problem.  We believe that this problem is not insurmountable, but the eventual solution might require a novel approach for treating div B.  As such, we have much work to do in improving the magnetohydrodynamics part of our code.  Another possible idea is to use the same basic numerical scheme, but impose the geometry of the cells by hand rather than determining them based on the motion of mesh points.  If such a scheme were developed, constrained transport could potentially be possible, since the geometry of the cells could be known at runtime.  This idea could have many other possible advantages, so it is a thought worth pursuing, but for TESS as it is currently written, we would like a better means for damping the growth of magnetic monopoles.

On the other hand, for some problems, div B may not be the most important concern.  If the magnetic divergence does not have time to grow, we may get to the end of a calculation before it starts to have an effect on the solution.  As an example, we show the relativistic MHD rotor test \citep{delzanna}, a challenging test of the robustness of our numerical scheme.  The calculation is done in a box of size ${[-.5,.5] \times [-.5,.5]}$.  The initial conditions are given by the following:

\begin{equation}
\rho = \left\{ \begin{array}
                  {l@{\quad:\quad}l}
                 10.0 & r < 0.1  \\  
                 1.0   & r > 0.1
                  \end{array} \right.
\end{equation}
\begin{equation}
P = 1.0, \qquad B_x = 1.0.
\end{equation}

The disc ${r<0.1}$ is initially rigidly rotating with angular frequency ${\omega = 9.95}$, so that the the edge of the disc has lorentz factor ${\gamma \approx 10}$.  The velocity field is zero outside the disc.  The calculation was done using a ${400 \times 400}$ mesh, and the results are shown in Fig. \label{fig:rotor} at time ${t=0.4}$.  This is ordinarily a very challenging RMHD test problem, but we required no special fail-safe procedures to solve it.  Actually, for this particular problem, we are at a slight disadvantage, since the mesh does not necessarily give us adaptive resolution exactly where we want it; the rarefaction is under-resolved (we start with an initially uniform mesh, but the central cells expand by roughly a factor of five in length).  However, it seems that, at least for our method and for this problem, div B does not interfere with our results.

\begin{figure}
\epsscale{1.0}
\plotone{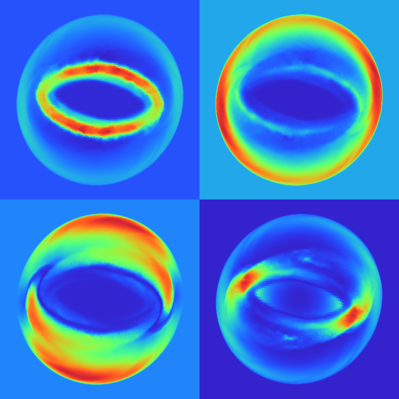}
\caption{ Relativistic MHD rotor test on a ${400 \times 400}$ mesh.  Upper left is density (${0.32 < \rho < 5.8}$), upper right is pressure (${p < 3.7}$), lower left is magnetic pressure (${.5 B^2 < 2.3}$), and lower right is Lorentz factor (${\gamma < 1.5}$).  TESS passes this test without requiring any fail-safe backup procedures.
\label{fig:rotor} }
\end{figure}

\subsection{Kelvin Helmholtz Instability}

\begin{figure*}
\epsscale{1.0}
\plotone{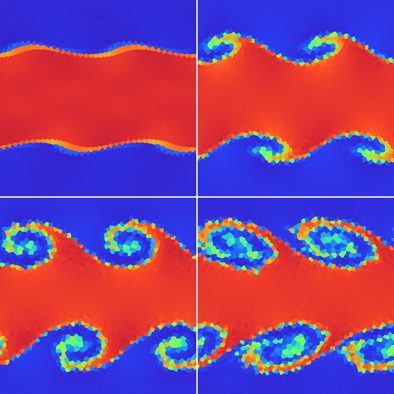}
\caption{ Nonrelativistic Kelvin Helmholtz instability on a 50x50 mesh, plotted at time intervals of ${t = 0.5, t = 1.0, t = 1.5,}$ and ${t = 2.0}$.  Color represents density.
\label{fig:kh1} }
\end{figure*}

\begin{figure}
\epsscale{1.0}
\plotone{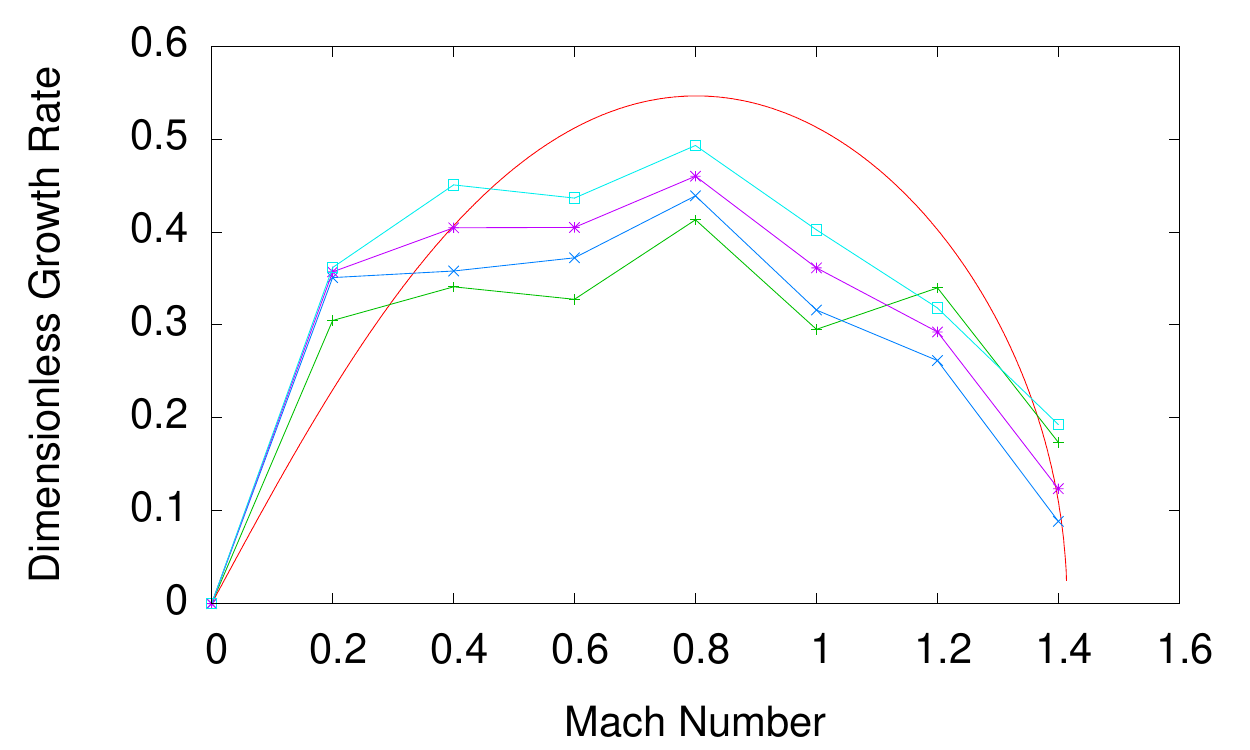}
\caption{ Linear growth rates of the relativistic Kelvin-Helmholtz instability, calculated using TESS at resolutions of 64, 128, 256, and 512 (squared).  The solid line is the analytical solution.
\label{fig:khgrowth} }
\end{figure}

\begin{figure*}
\epsscale{1.0}
\plotone{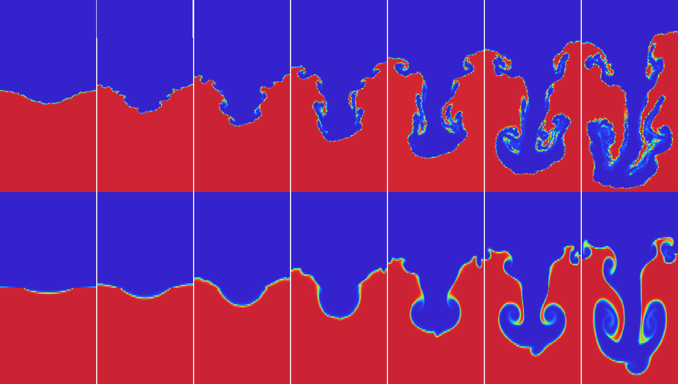}
\caption{ Relativistic Rayleigh-Taylor instability, at a resolution of 128$\times$128, with a relativistic Atwood number of ${\mathcal{A} = 0.6}$, for a moving mesh (above) and fixed mesh (below), plotted at regular intervals until ${t = 7.0}$.  Asymmetry is primarily the result of small perturbations initially given to the mesh generators.  These same small perturbations were given in the fixed mesh case, so that the only difference between the two versions is whether the mesh points are moving.
\label{fig:rt1} }
\end{figure*}

We find that TESS is very good at solving problems involving fluid instabilities, particularly in cases where a contact discontinuity is being perturbed.  The Kelvin-Helmholtz instability is one such example, which occurs at the interface between shearing fluids.  We look at the nonrelativistic case first, to compare with AREPO.

We use a periodic box, ${[0,1] \times [0,1]}$, with a stripe in the middle half of the box ${1/4 < y < 3/4}$.  The pressure is uniform throughout the domain, and the density and x-velocity are uniform in each region:

\begin{equation}
P = 2.5
\end{equation}
\begin{equation}
\rho =  \left\{ \begin{array}
                  {l@{\quad:\quad}l}
                 2 & .25 < y < .75  \\  
                 1 & otherwise
                  \end{array} \right.
\end{equation}
\begin{equation}
v_x =  \left\{ \begin{array}
                  {l@{\quad:\quad}l}
                 .5 & .25 < y < .75  \\  
                 -.5 & otherwise
                  \end{array} \right.
\end{equation}

This sets up the initial conditions for the instability, and to excite a particular mode we add a small perturbation to the y-component of the velocity:

\begin{equation}
v_y = w_0 sin(4 \pi x) f(y)
\end{equation}
\begin{equation}
f(y) = exp(-{ (y-.25)^2 \over 2 \sigma^2}) + exp(-{(y-.75)^2\over 2 \sigma^2}) .
\end{equation}

We choose ${w_0 = .1}$ and ${\sigma = .05/\sqrt{2}}$.  We use an adiabatic index of ${\Gamma = 5/3}$.  These initial conditions are identical to those used by \cite{arepo}.  Our visualization is different, so it is difficult to directly compare results, but it appears that the growth rate of the instability is the same for a 50x50 lattice (see Fig. \ref{fig:kh1}).

\begin{figure}
\epsscale{1.0}
\plotone{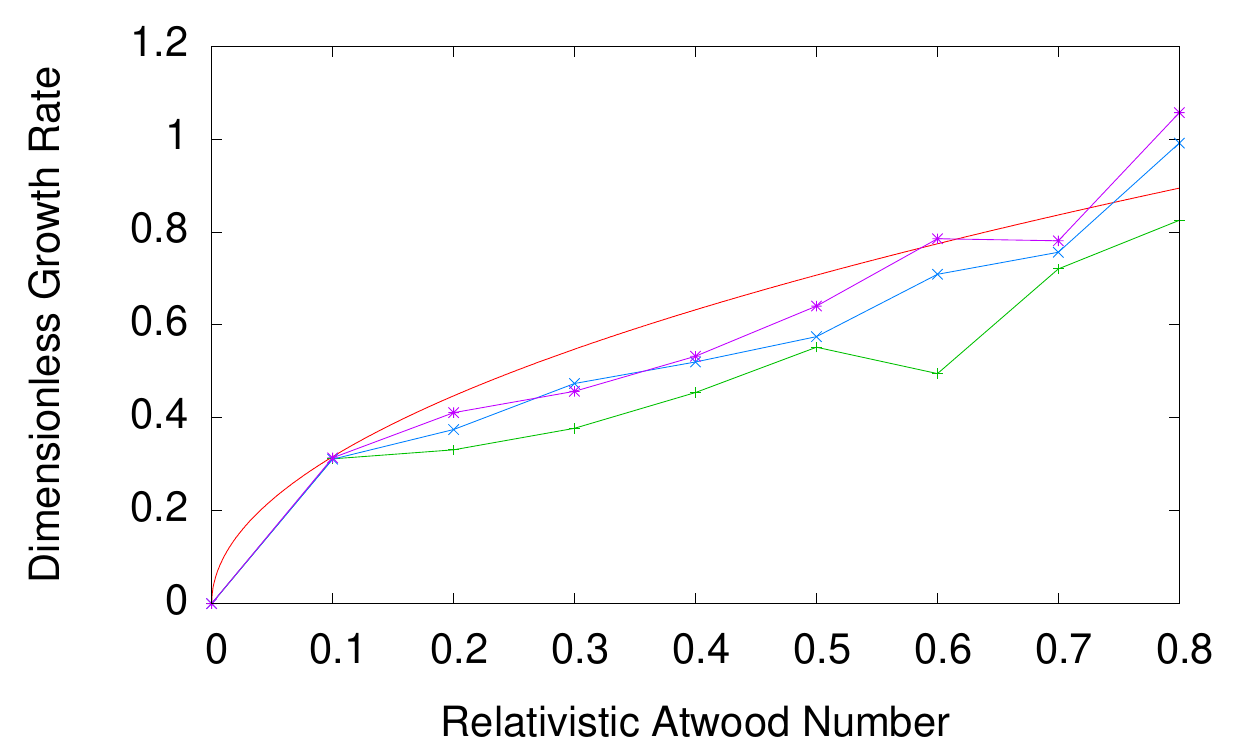}
\caption{ Linear growth rates of the relativistic Rayleigh-Taylor instability, calculated using TESS at resolutions of 64, 128, 256, and 512 (squared).  The solid line is the analytical solution.
\label{fig:rtgrowth} }
\end{figure}

Tests of Galilean invariance, on the other hand, indicate that our numerics do depend on the moving reference frame in which the calculation is performed.  When adding an overall velocity to the initial conditions, we find that the results are significantly changed if the boost velocity is large compared to the flow velocity.  This can be traced back to the fact that the HLLC solver is not invariant under Galilean boosts.  It would be interesting to see how differently the code would perform using an exact Riemann solver, though this is not a very high priority, because TESS is ultimately designed to solve relativistic problems.  A more optimistic way to interpret this result is to say that Galilean invariance is not a necessary condition for accurately capturing contact discontinuities.

As a relativistic test, we use TESS to calculate the growth rate of the Kelvin-Helmholtz instability in the relativistic case.  For this purpose, we change the initial conditions slightly:  the initial density is uniform (${\rho = 1}$) everywhere, the pressure is now ${P = 1}$ and the adiabatic index ${\Gamma = 1.4}$.  The initial perturbation is given to the y component of the four-velocity, ${u^y}$, and the x component ${u^x}$ is varied.  To measure the development of the instability, we add an additional passive scalar quantity, X, which is advected with the fluid velocity, for which we choose ${X=0}$ outside the stripe and ${X=1}$ inside the stripe.  We measure the growth of the instability by tracking the position of the contact discontinuity in X.  Note that this is much more straightforward to do with TESS than it would be for an Eulerian code.  In fact, we can track the development of the instability even while the perturbation to the contact is smaller than the size of a Voronoi cell.

We compare the linear growth rate against the analytic solution.  The growth rate has been calculated analytically by \cite{kh}.  It is the imaginary part of ${\phi k c_s}$ where $k$ is the wavevector corresponding to the wavelength of the perturbation (${k = 4\pi}$ in this case), ${c_s}$ is the sound speed, and ${\phi}$ is given by the following formula:

\begin{figure*}
\epsscale{1.0}
\plotone{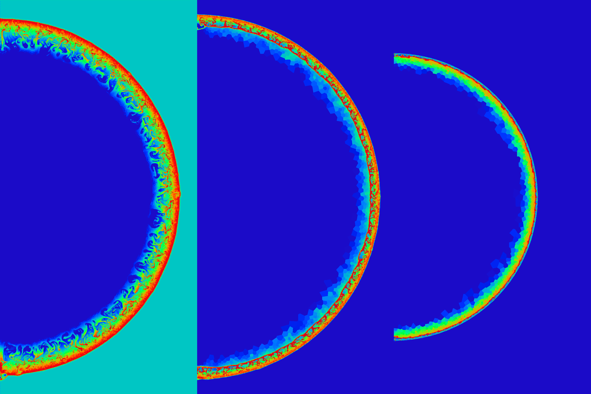}
\caption{ Relativistic Rayleigh-Taylor instability in a decelerating shock.  The initial density profile is given by a power law, with k = (left to right) 0, 1, and 2.  In spherical symmetry the initial conditions would result in a transient two-shock solution, but in more than one dimension, the contact discontinuity is unstable and it is disrupted enough to generate unstable motion all throughout the forward shock.  The first two calculations have been terminated at ${t=.95}$ when the shock has decelerated to ${\gamma \sim 4}$, the last one at ${t=.7}$, while it is still highly relativistic, ${\gamma \sim 11}$.
\label{fig:rtx1} }
\end{figure*}

\begin{equation}
{\phi^2 \over \mathcal{M}^2} = {1-v^2 \over 1-c_s^2} { {\mathcal{M}^2 + 1 - v^2 - \sqrt{ 4 \mathcal{M}^2(1-v^2) + (1+v^2)^2 } } \over {\mathcal{M}^2 + 2 v^2 } }
\end{equation}
Here, v is the flow speed, and ${\mathcal{M}}$ is the relativistic Mach number, defined:
\begin{equation}
\mathcal{M} = {\gamma v \over \gamma_s c_s}
\end{equation}
where ${c_s}$ is the relativistic sound speed, and ${\gamma_s}$ is the corresponding Lorentz factor.

In Fig. \ref{fig:khgrowth} we plot the numerically calculated growth rates against this analytical formula.  At ${64 \times 64}$ resolution, the growth rate was not very accurately calculated, but for all other tests we seem to get very accurate results, considering the low resolution.

Notice that in contrast with the nonrelativistic example, it does not make sense to even ask about the Lorentz invariance of the code for this problem, since periodic boundary conditions are only periodic in one reference frame, due to relative simultaneity.  In general, even if we do not employ periodic boundaries, testing Lorentz invariance can be an extremely complicated task, because one would have to take into account that different mesh points can be sampling the fluid motion at different points in time.  In any case, we know that our code is not manifestly Lorentz invariant, so we would not expect a striking improvement in Lorentz invariance over an Eulerian code.

\subsection{Rayleigh Taylor Instability}

\begin{figure*}
\epsscale{1.0}
\plotone{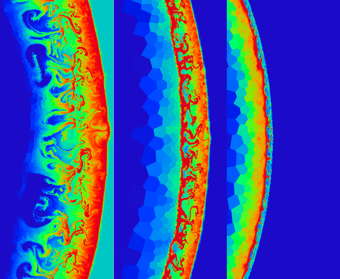}
\caption{ A close-up of Figure \ref{fig:rtx1}.
\label{fig:rtx2}}
\end{figure*}

A related fluid instability is the Rayleigh-Taylor instability, where a fluid of high density is balanced on top of a fluid of low density, and the contact discontinuity between these two different densities is perturbed by the force of gravity.  In the relativistic case, we can get a gravitational field by transforming to an accelerated reference frame.  This is not difficult to do, but we can also simply solve the relativistic equations in the weak-field limit.  For the purposes of calculating the linear growth rate, it would not matter which we do, since for a small enough perturbation, the potential jump between the highest and lowest points of the perturbation will be small, and hence the weak-field limit applies.  We set up the following initial conditions in the computational domain ${(x,y) \in [0,1] \times [-1,1]}$:

\begin{equation}
\rho = \left\{ \begin{array}
                  {l@{\quad:\quad}l}
                 1.0 & y < 0  \\  
                 \rho_2 & y > 0
                  \end{array} \right.
\end{equation}
\begin{equation}
P = P_0 e^{-gy/(\Gamma-1)} +( \Gamma-1)\rho(e^{-gy/(\Gamma-1)} - 1)
\end{equation}
For the initial velocity perturbation:
\begin{equation}
v_y = w_0 cos(2 \pi x) exp(-{y^2 \over 2 \sigma^2} ) .
\end{equation}

We choose constants ${P_0 = 10}$, ${w_0 = 0.03}$, ${\sigma = .05/\sqrt{2}}$, and the gravitational field is ${g = 0.1}$.  ${\rho_2}$ is varied to get different growth rates.

In Fig. \ref{fig:rt1}, we show the growth of the instability using a moving or fixed mesh, at fairly low resolution, 128x128.  The initial mesh was not perfectly square; the cells were randomly offset slightly to improve the regularity of the mesh throughout the calculation.  This small (~.1\%) offset is enough to cause asymmetry, even in the fixed-mesh case, but we decided to employ the offset in both cases, so that the only difference between the two cases is whether the mesh points are allowed to move.

The linear growth rate of the instability is easy to derive even in the relativistic case, because the only relativistic effect for small perturbations is that gravity couples to energy instead of mass.  Therefore, the growth rate for the relativistic case is still
\begin{equation}
R = \sqrt{ g k \mathcal{A} },
\end{equation}
Where $g$ is the gravitational field strength, $k$ is the wavevector corresponding to the perturbation (${k=2\pi}$ in this case), and ${\mathcal{A}}$ is the relativistic Atwood number, defined as
\begin{equation}
\mathcal{A} = { \rho_2( 1+e_2 ) - \rho_1( 1+e_1 ) \over \rho_2 ( 1 + e_2 ) + \rho_1 ( 1 + e_1 ) }
\label{eqn:rtgrowth}
\end{equation}

In Fig. \ref{fig:rtgrowth} we plot the growth rates numerically calculated using TESS, and compare with the analytical result (\ref{eqn:rtgrowth}).  Even at extremely low resolutions, we capture the growth rates quite accurately.

\section{Rayleigh Taylor Instability in a Decelerating Shock}

The Rayleigh-Taylor instability provides us with a useful astrophysical application, in the context of a decelerating relativistic shock.  The reason this is relevant is that an external gravitational field is equivalent to an accelerating reference frame.  Hence a low-density gas accelerating into a high-density gas or a high-density gas decelerating into a low-density gas will exhibit this same instability.  The latter could potentially occur in a shockwave, as matter is pushing its way through some ambient medium \citep{levinson}.  If the explosion is spherical, and the density profile has a power law dependence, ${ \rho \sim r^{-k} }$, then the shock will decelerate for ${k < 3}$.  If we assume a simple point explosion in this density profile, we recover the Blandford-McKee solution \citep{bmk}, or at late enough times, the nonrelativistic Sedov-Taylor solution.  In both cases, the shock front and contact discontinuity coincide.  However, if there is excess mass in the initial explosion, so that the total mass is greater than the integral of the power-law density profile, then there will be a coasting period until the shock overtakes an amount of mass comparable to this excess mass.  This is followed by deceleration to a two-shock solution \citep{ns06}, because the information that the shock is now decelerating needs to be transported back upstream.  In this case, there will be a contact discontinuity between the two shocks, and the density discontinuity across this contact can be quite significant.  For this situation, the Rayleigh-Taylor instability can play an important role in the solution, because the contact is quite unstable.  A calculation in spherical symmetry would be incorrect, even though the initial conditions are spherically symmetric.

We capture the Rayleigh-Taylor instability from a spherically symmetric explosion in two dimensions using axisymmetric coordinates, for density profiles corresponding to ${k=0,1,2}$.  We use the domain ${(s,z) \in [0,1] \times [-1,1]}$ where ${r^2 = s^2 + z^2}$ and set up the initial conditions as follows:
\begin{equation}
\rho = \left\{ \begin{array}
                  {l@{\quad:\quad}l}
                 \rho_0 (r_0/r_{min})^{k_0} & r<r_{min} \\
                 \rho_0 (r_0/r)^{k_0} & r < r_0 \\  
                 \rho_0 (r_0/r)^{k} & r > r_0 \\
                  \end{array} \right.
\end{equation}
We set ${k_0 = 4}$ and ${r_0 = .1}$ to start with an accelerating shock which typically coasts until around ${r \sim .2}$, then decelerates.  We use ${r_{min} = .001}$ to ensure that the system has a finite amount of mass.  The initial pressure profile is given by
\begin{equation}
P = \rho \left\{ \begin{array}
                  {l@{\quad:\quad}l}
                 e_0/3 & r < r_{exp} \\  
                 10^{-6} & r > r_{exp} \\
                  \end{array} \right.
\end{equation}

We use a gamma-law equation of state with ${\Gamma = 4/3}$.  We choose ${r_{exp} = .003}$, and choose ${e_0}$ to get a shock which reaches Lorentz factors of ${\gamma \sim 10}$ before it decelerates to mildly relativistic flow by ${t = .95}$, when the calculation ends.  The values of ${e_0}$ are given in table \ref{tab:e0}, along with a rough estimate of the bulk Lorentz factor for the fluid at its maximum speed, and also the Lorentz factor at the end of the calculation.  Note that these initial conditions are fundamentally different from the conditions chosen by \cite{levinson}; his results came from perturbative analysis of the self-similar solution of \cite{ns06}, whereas ours begins with an explosion.  In spherical symmetry, our explosion approaches this self-similar solution, but due to the instability, in 2D our calculation should never find this solution (at late enough times, however, it should eventually find the Sedov solution).

\begin{figure*}
\epsscale{1.0}
\plotone{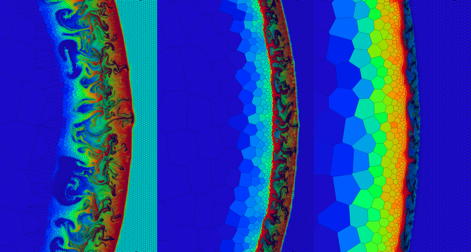}
\caption{ The final state of the tessellation for the Rayleigh-Taylor explosion.  Panels are the same as in Figures \ref{fig:rtx1} and \ref{fig:rtx2}.  Voronoi cells have been compressed into the shock front, so that nearly all the resolution is concentrated there.
\label{fig:rtx3} }
\end{figure*}

\begin{deluxetable}{cccc}
\tablecaption{Rayleigh-Taylor explosion parameters
\label{tab:e0}}
\tablewidth{0pt}
\tablehead{
\colhead{Power Law (k)}  &
\colhead{${e_0}$} &
\colhead{${\gamma_{max}}$} & 
\colhead{${\gamma_{final}}$}
}
\startdata
0 & 6 & 11 & 4 \\
1 & 4 & 9 & 4 \\
2 & 6 & 13 & 11
\enddata
\end{deluxetable}

In Figures \ref{fig:rtx1} - \ref{fig:rtx3} we show the growth of the instability for three power-law density profiles.  It appears that the instability grows rapidly enough to catch up to the forward shock.  In fact, relativity appears to help us in this case, because for large Lorentz factors the shock front must remain a short distance from the contact discontinuity, keeping it within reach until the Rayleigh-Taylor fingers can catch up to the shock.  The fact that the instability overtakes the forward shock appears to be generic and does not require special initial conditions, although it would of course not occur for a simple point explosion, because in this case the shock and contact discontinuity would already coincide.  Using steep power-laws with little deceleration does not seem to change this picture; the instability can still catch up to the shock even when ${k=2}$ and the lorentz factor is 11, as in our third test case.  This result is in qualitative agreement with the results of \cite{levinson}, whose work was restricted to the linear growth rate.  To our knowledge, ours is the first direct numerical calculation of the instability for a relativistic shock.

It is worth noting another advantage to TESS which is shown most plainly in Fig. \ref{fig:rtx3}.  If we know in advance where to initially place the cells, we can get high resolution exactly where we need it.  In this case, we initially concentrated most of the cells near the origin, and the region of high resolution followed the motion of the shock.  Nearly all of the zones have been compressed into the shock front, giving us very high resolution there.  We generated this initial mesh in the following way:  We started with an initial ${512 \times 1024}$ uniform mesh in the domain ${(x,y) \in [0,1] \times [-1,1]}$ (though we staggered mesh points so that our initial mesh was not exactly square).  We then changed the location of the mesh points via the following prescription:
\begin{eqnarray}
r = max( |x| , |y| ) \\
x \rightarrow x e^{k(r-1)} \\
y \rightarrow y e^{k(r-1)}
\end{eqnarray}
where we chose ${k = 3.5}$.  Thus our initial resolution was ${e^k \sim 33}$ times higher near the origin than it would be for a uniform mesh.  It should be noted that this effective resolution changes throughout the time evolution, partially due to the shock overtaking more cells, and partially due to the increasing amount of volume which is being well resolved.

\section{Summary}
\label{sec:summary}

TESS is a new relativistic hydrodynamics code based on the Voronoi
tessellation.  It performs particularly well on problems with sharp
discontinuities, especially contact discontinuities.  TESS has been
extensively tested. For smooth problems, convergence is slightly
better than second order.  For problems involving a moving contact
discontinuity, TESS demonstrated a clear advantage over the Eulerian
codes in that smearing of contact discontinuities is greatly reduced.
It was also demonstrated that employing the HLLC solver was necessary
to get full advantage out of the Lagrangian nature of this method.
For many nonrelativistic problems, TESS has success comparable to that
of AREPO, but unfortunately lacks the Galilean invariance property.

We have applied TESS to an astrophysical example, testing the
stability of a decelerating spherical shock wave.  We found that the
contact discontinuity behind the shock was unstable to the
Rayleigh-Taylor instability, so much so that the instability was able
to reach the forward shock.  A detailed study will be presented in a
future publication.

The structure of TESS is relatively simple, there is no need to
perform any explicit rotations or boosts, and we therefore have the
luxury of using an approximate Riemann solver, so long as the solver
doesn't inherently diffuse contact discontinuities.  This is a
definite advantage, because for some hyperbolic systems, calculating
the exact solution to the Riemann problem is computationally intensive
if not impossible.  Our method can be used to solve a wide range of
hyperbolic problems, like general relativistic hydrodynamics, or
magnetohydrodynamics.  The tessellation algorithm is also quite
simple, and very robust; it takes little effort to implement, and is
not bug-prone.

In speed tests, we find the tessellation algorithm consumes roughly half of the code's run-time.  We urge care in the interpretation of this, however, as very little work has been spent in optimization, and moreover this figure is highly problem-dependent, as is the code's overall runtime.  On a 2.67GHz Intel Xeon X5550 processor, TESS spends roughly 40 microseconds per zone per timestep, which is comparable to RAM's efficiency.  We hope to improve the efficiency in future developments.

The code is structured in such a way that an extension to three dimensions merely requires writing a three-dimensional tessellation algorithm, which would be an extension of the two-dimensional algorithm already outlined.  Making the code run in parallel does not pose any major academic hurdles (apart from load-balancing).  Both of these ideas are extensions to the code which we are currently pursuing.  We are also interested in implementing more complicated boundary conditions, mesh refinement, and a local time step, as these have all been implemented in AREPO and are thus a solved problem.

\acknowledgments
This work was supported in part by NASA under grant NNX10AF62G issued through the Astrophysics Theory Program (ATP) and by the NSF through grant
AST-1009863.  The authors would like to thank the Institute for Theory and Computation at the Harvard-Smithsonian Center for Astrophysics for hospitality while finishing this work.  We are grateful to John Hawley, Debora Sijacki, Volker Springel, Jim Stone, Mark Vogelsberger, Weiqun Zhang, and Jonathan Zrake for helpful comments or discussions.  We would also like to thank the anonymous referee for his/her thorough review.

\begin{appendix}
\section{Primitive Variable Solver for Relativistic Magnetohydrodynamics}
\label{app}

The conversion from conserved variables back to their primitive counterparts is nontrivial for relativistic magnetohydrodynamics.  It is particularly challenging if we wish to use an arbitrary equation of state.  Our code employs a three-dimensional Newton-Raphson solver to recover the primitive variables from the conserved variables.  The equations being solved are nearly the same as those given by \cite{cons2prim}; we derive them again here for completeness.

We begin with the eight conserved variables, which we know:
\begin{equation}
D = \gamma \rho, \qquad \vec B, \qquad Q^\mu = T^{0 \mu}_{fluid} + T^{0 \mu}_{EM}
\end{equation}
We also make the following convenient definitions:
\begin{equation}
K \equiv \gamma^2 v^2 = \gamma^2-1
\end{equation}
\begin{equation}
W \equiv \gamma^2 \rho h = \sqrt{K+1} D + (K+1)( \epsilon + P )
\label{eqn:W}
\end{equation}
The electromagnetic energy and momentum is straightforward to calculate assuming ${\vec E = - \vec v \times \vec B}$:
\begin{eqnarray}
T^{0 0}_{EM} = {1 \over 2} ( E^2 + B^2 ) = {1 \over 2} ( (v \times B )^2 + B^2 ) = B^2 - {1 \over 2 } ( B^2 + (\vec u \cdot \vec B)^2 ) / \gamma^2 \\
T^{0 i}_{EM} = ( E \times B )^i = - ( ( v \times B ) \times B )^i = B^2 v^i - B^i (\vec u \cdot \vec B)/ \gamma
\end{eqnarray}
So, the four-momentum Q can be written out explicitly:
\begin{eqnarray}
Q^0 = W - P + B^2 -  {1 \over 2} { B^2 + (\vec u \cdot \vec B)^2 \over \gamma^2}  \\
\vec Q = ( W + B^2 ) \vec v - ( \vec u \cdot \vec B) \vec B /\gamma
\end{eqnarray}
Following Noble, we can also take the dot product ${\vec Q \cdot \vec B}$ to evaluate ${\vec u \cdot \vec B}$:
\begin{equation}
\vec u \cdot \vec B = \gamma \vec B \cdot \vec Q / W
\end{equation}
If we substitute this back into the equations for the four-momentum, we arrive at the following two relations:
\begin{equation}
Q^0 = W - P + B^2 -  {1 \over 2}(  B^2 / (K+1) + ( \vec B \cdot \vec Q / W )^2  )
\label{eqn:c2p1}
\end{equation}
\begin{equation}
\vec Q^2 = ( W + B^2 )^2 K/(K+1)  - (2 W + B^2) ( \vec B \cdot \vec Q)^2 / W^2
\label{eqn:c2p2}
\end{equation}
Equations (\ref{eqn:c2p1}) and (\ref{eqn:c2p2}), along with the definition for W (\ref{eqn:W}), constitute three equations with the three unknown variables, W, K, and the temperature T.  We can now easily turn this into a Newton-Raphson rootfinding scheme where we search the three-dimensional parameter space of W,K, and T for a solution to the three equations.  All of this assumes an arbitrary equation of state,
\begin{eqnarray}
P = P(\rho,T) = P( {D \over \sqrt{K+1}} , T ) \\
\epsilon = \epsilon (\rho,T) = \epsilon( {D \over \sqrt{K+1}} , T )
\end{eqnarray}
Note that ${W, K , T \in [0,\infty)}$, so that we don't have to worry about what to do if one of our variables is too large (this would be the case if, for example, we used velocity as one of our unknowns).  If K becomes negative on a given iteration, it is reset to zero to prevent taking a square root of a negative, but W and T are generally not constrained.  For an adiabatic equation of state, we use the same framework but make the definition
\begin{equation}
P(\rho,T) \equiv T, \qquad \epsilon(\rho,T) \equiv T/(\Gamma - 1)
\end{equation}
so that effectively pressure takes the place of temperature as the third variable being solved for.  For our initial guess values, we use the values of W,K,and T from the previous timestep.  However, if the solver fails to find a root, we try again using new guess values.  One idea is to try an estimate which is exact in the limit that rest mass or magnetic energy dominates.  In doing so, we obtain the following estimates for K:
\begin{equation}
K_D = { \vec Q^2 \over D^2 } \qquad K_B = { (\vec Q \cdot \vec B)^2 \over D^2 B^2 }
\end{equation}
To take into account both of these possibilities, we use a guess value which is a weighted average of these two estimates:
\begin{equation}
K_{est} = { D K_D + B^2 K_B \over D + B^2 }
\end{equation}
If this estimate fails, we can use a method found by \cite{cd08} and try a more safe set of guess values, based on the maximum possible values (also assuming the Lorentz factor is never larger than ${10^4}$):
\begin{equation}
\epsilon^{*} = ( Q^0 - D - B^2/2 )/D, \qquad \rho^{*} = D , \qquad P^{*} = P( \rho^{*} , \epsilon^{*} )
\end{equation}
\begin{equation}
W_{guess} = Q^0 + P^{*} - B^2/2 , \qquad K_{guess} = 10^8 , \qquad T_{guess} = T( \rho^{*} , \epsilon^{*} )
\end{equation}
To determine when we have converged on the inverted state, we need some error function which will measure how close we are to the true solution.  For example, one natural choice is
\begin{equation}
\delta = | {\Delta W \over W }|
\end{equation}
where ${\Delta W}$ measures how much W has changed since the last iteration.  The strategy then would be to iterate until ${\delta < TOL}$ for some specified tolerance, typically ${TOL \sim 10^{-11}}$.  Unfortunately, it's possible for W to be changing very slowly while K and T are still far from the solution.  To prevent this from being a problem, we modify the error function as follows:
\begin{equation}
\delta =  | {\Delta W \over W }| +  ({\Delta K \over K })^2 +  ({\Delta T \over T })^2
\end{equation}
This agrees with the previous definition when we are close to the true solution, but away from the solution, it helps to prevent "false positives", where ${\delta}$ becomes small before the true solution has been found.

\end{appendix}

{}

\end{document}